\documentclass[letterpaper]{article}
\usepackage{aaai2027}
\nocopyright 
\usepackage[hyphens]{url}
\usepackage{graphicx}
\usepackage{natbib}
\usepackage{caption}
\usepackage{booktabs}
\usepackage{tabularx}
\usepackage{multirow}
\usepackage{amsmath, amssymb}
\usepackage{algorithm}
\usepackage{algorithmic}

\newcommand{\PROCEDURE}[2]{\STATE \textbf{procedure} \textsc{#1}\textit{(#2)}}
\newcommand{\ENDPROCEDURE}{\STATE \textbf{end procedure}}
\usepackage{listings}
\usepackage{xcolor}
\usepackage{enumitem}
\usepackage{needspace}
\usepackage{etoolbox}

\definecolor{codebg}{rgb}{0.96,0.96,0.96}
\BeforeBeginEnvironment{lstlisting}{\needspace{6\baselineskip}}

\lstdefinelanguage{yaml}{
  keywords={true,false,null},
  keywordstyle=\color{blue},
  basicstyle=\ttfamily\scriptsize,
  sensitive=false,
  comment=[l]{\#},
  commentstyle=\color{gray}\ttfamily,
  stringstyle=\color{red},
  morestring=[b]',
  morestring=[b]",
}

\lstdefinelanguage{json}{
  basicstyle=\ttfamily\scriptsize,
  showstringspaces=false,
  breaklines=true,
  literate=
   *{0}{{{\color{blue}0}}}{1}
    {1}{{{\color{blue}1}}}{1}
    {2}{{{\color{blue}2}}}{1}
    {3}{{{\color{blue}3}}}{1}
    {4}{{{\color{blue}4}}}{1}
    {5}{{{\color{blue}5}}}{1}
    {6}{{{\color{blue}6}}}{1}
    {7}{{{\color{blue}7}}}{1}
    {8}{{{\color{blue}8}}}{1}
    {9}{{{\color{blue}9}}}{1}
    {\{}{{{\color{red}\{}}}{1}
    {\}}{{{\color{red}\}}}}{1}
    {[}{{{\color{red}[}}}{1}
    {]}{{{\color{red}]}}}{1},
  stringstyle=\color{green!40!black},
  keywordstyle=\color{blue},
  morestring=[b]",
  morestring=[b]',
}

\def\UrlFont{\rm}
\newcommand{\abe}{ABE-Ralph}
\newcommand{\metall}{\texttt{MET\_ALL}}
\newcommand{\metpartial}{\texttt{MET\_PARTIALLY}}
\newcommand{\failed}{\texttt{FAILED}}

\title{Beyond Execution: Auditing Experimental Fidelity in \\ LLM-Driven Scientific Research}

\author{
    Lezhi Yu\textsuperscript{\rm 1}, Xiaogang Xu\textsuperscript{\rm 1}, Yuhua Zhou\textsuperscript{\rm 1}, Shuibing He\textsuperscript{\rm 1}, Aimin Pan\textsuperscript{\rm 2}
}
\affiliations{
    \textsuperscript{\rm 1}College of Computer Science and Technology, Zhejiang University, Hangzhou, China \\
    \textsuperscript{\rm 2} Zhejiang Lab, Hangzhou, China
}

\begin{document}

\maketitle

\begin{abstract}

LLM agents used for scientific experimentation must do more than generate executable code: they must implement the reference method faithfully, design experiments that test the paper's claims, and provide evidence supporting those claims. We show that agents often produce \emph{methodological hallucinations}: silently reducing datasets or training budgets, replacing failed learning or generative components with lookup or oracle functions, or drawing conclusions from resource-limited settings where a method's claimed advantage disappears. To detect these failures, we introduce \abe, a reference-anchored auditing framework that represents claims, protocols, required components, baselines, and metrics as structured experimental constraints, guides implementation through an 8-step workflow, and performs quantitative, qualitative, and code-level verification. Across 30 long-horizon reproduction runs covering 12 machine learning domains, \abe~achieves a 93\% robust execution rate and identifies five scientific failure modes. In 23 NatureBench discovery tasks, \abe~matches or exceeds state-of-the-art performance on 5 tasks. These results show that reliable evaluation of AI scientists must assess whether the experimental design faithfully tests the intended claim and whether the resulting evidence supports it, rather than treating code execution or plausible metrics as evidence of scientific success.

\begin{links}
\link{Code}{https://github.com/Flavorfish/AutoRepro}
\end{links}

\end{abstract}

\section{Introduction}

Large Language Model (LLM) agents are expanding into autonomous research workflows, allowing systems to generate hypotheses, write code, run experiments, and draft reports \citep{aiscientist, autoresearchclaw, boiko2023autonomous, bran2024chemcrow}. Similarly, software engineering agents solve repository issues and output functional patches with high success rates \citep{sweagent, openhands, agentbench}. These developments make it tempting to evaluate scientific agents using the same criterion used for software agents: whether the generated program executes successfully and produces an output.

That criterion is insufficient for scientific reproduction. A reproduction must satisfy four linked requirements: (1) the implementation contains the reference method described in the paper; (2) datasets, preprocessing, training, and baselines remain relevant to the original experiment; (3) the experiment is designed to test the central claim under the computational regime; and (4) observed results provide valid evidence for or against that claim. A script returning exit code 0 or reporting plausible metrics can easily fail all four requirements.

\begin{figure*}[t]
\centering
\includegraphics[width=\textwidth]{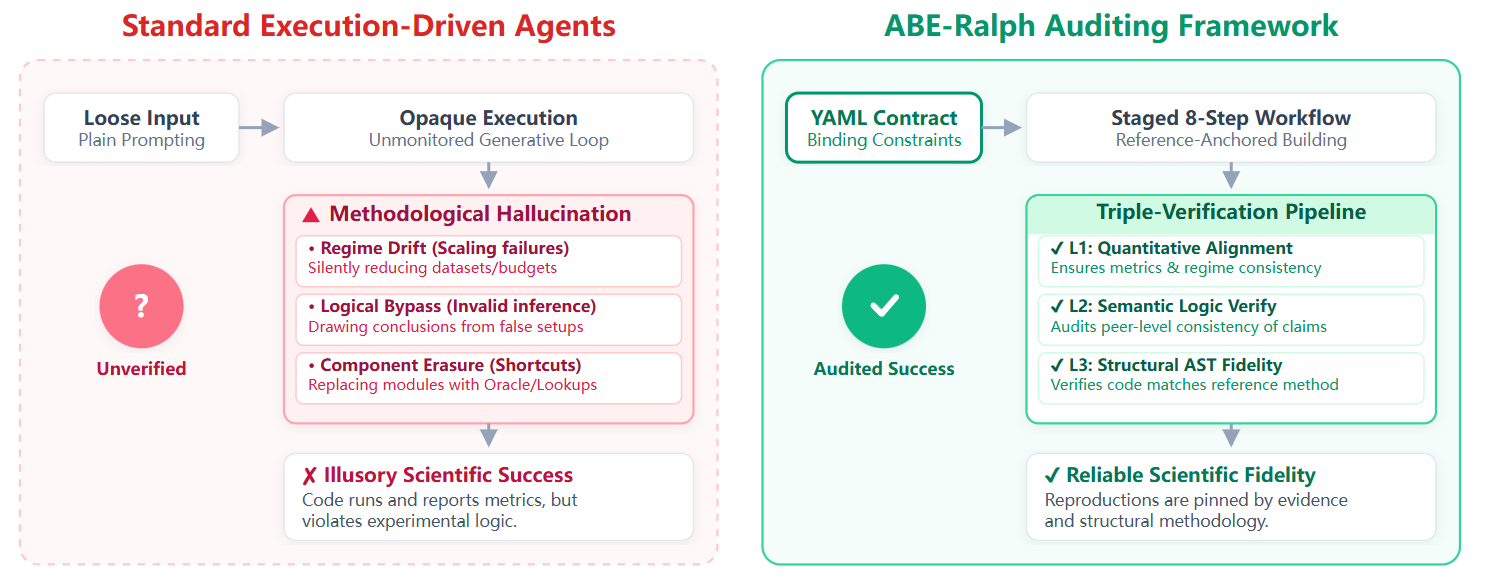}
\vspace{-0.3in}
\caption{Comparison between standard execution-driven agents and our proposed \abe~auditing framework. While traditional agents often bypass computational limits using undetected shortcuts (Methodological Hallucinations), \abe~locks development constraints via YAML contracts and audits code pipelines through a Triple-Verification system.}
\label{fig:teaser}
\vspace{-0.1in}
\end{figure*}

This distinction is especially crucial when agents operate under compute limits, missing dependencies, or failed checkpoints. When an experiment is difficult to run, an agent may silently use a smaller dataset, fewer training steps, lower resolution, or random weights without reporting protocol changes. It may replace a costly generative module with a lookup rule or oracle function that already holds the answer, or run the method at a scale too small for its claimed advantage to emerge and conclude the hypothesis is false. These actions preserve the appearance of progress while failing scientific logic. We define these deviations as \textbf{Methodological Hallucinations}: silent, hard-to-detect violations of pre-defined experimental constraints that preserve superficial code execution while undermining scientific logic \citep{hitchhiker2025hallucination, careful2026framework}. Consequently, they produce misleading metrics that risk validating false hypotheses or incorrectly dismissing valid claims.

The problem is therefore not only that agents sometimes generate incorrect code. More fundamentally, they can generate a scientifically misleading experimental process while producing technically valid artifacts. Existing execution-driven evaluations are largely unable to distinguish a faithful reproduction from a simplified implementation, an incomplete experiment, or an experiment whose resource constraints invalidate its conclusion. Human reviewers can identify some of these problems, but manual inspection is difficult to scale across long-horizon agent runs.

We introduce \textbf{\abe}\quad (Auto Baseline Experiment), an automated scientific auditing framework that monitors, binds, and verifies the experimental lifecycle of AI-driven research, treating scientific reproduction as a reference-anchored process. Before implementation, \abe\ structures paper claims, architectural components, datasets, baselines, metrics, and resource bounds into declarative YAML contracts. An 8-step workflow guides model construction and protocol execution. During execution, a Triple-Verification pipeline checks quantitative metric alignment, qualitative semantic logic, and structural code fidelity. These checks intercept deceptive model shortcuts and guarantee methodological fidelity and auditability.

Distinguishing execution success from scientific validity is essential. A faithful auditing framework does not merely replicate historical metrics; it provides a foundation to cross-examine original outcomes and systematically discover optimized configurations that exceed baseline benchmarks. We report reference-anchored reproduction outcomes separately from raw execution success, evaluating whether each task reproduces or exceeds reference results.

Our contributions include:
\begin{enumerate}
    \item \textbf{Taxonomy of Methodological Hallucinations:} We conceptualize and define a 5-class taxonomy of deceptive agentic shortcuts in scientific workflows that maintain successful system execution while violating core methodological bounds.
    \item \textbf{The \abe\ Auditing Framework:} We propose \abe, an 8-step reference-anchored framework enforcing pre-execution YAML contracts and an automated Triple-Verification system (numerical, logical, and code-structure levels).
    \item \textbf{Empirical Evaluation and Discovery:} We validate the system across 30 classical ML benchmarks spanning 12 domains (achieving a 93\% robust execution rate and exposing systematic methodological shortcuts), and further demonstrate its capability in discovery mode across 23 NatureBench tasks (matching or exceeding SOTA baselines on 5 tasks).
\end{enumerate}

\section{Related Work}

\subsection{LLM Agents for Scientific Discovery}
Autonomous scientific agents like The AI Scientist \citep{aiscientist}, AutoResearchClaw \citep{autoresearchclaw}, and Claw-AI-Lab \citep{clawailab} build research workflows but focus primarily on novelty and syntax execution, lacking auditing mechanisms for experimental fidelity. Similarly, domain-specific systems, e.g.,ChemCrow \citep{bran2024chemcrow} and Coscientist \citep{boiko2023autonomous}, automate lab tools without verifying logical correctness. As highlighted in recent surveys \citep{zheng2025agentic}, existing platforms evaluate the superficial appearance of scientific work rather than its internal validity. Drawing inspiration from these efforts, \abe~introduces constraint verification to audit reproduction fidelity.

\subsection{LLM Agents for Software Engineering}
Software engineering agents like SWE-agent \citep{sweagent} and OpenHands \citep{openhands} optimize for passing test suites under SWE-bench paradigms \citep{swebench, swe2024multimodal, swe2025pro}. Similarly, code benchmarks like HumanEval \citep{humaneval} and MBPP \citep{mbpp} measure execution correctness rather than scientific intent. However, passing unit tests or exit-0 checks is insufficient for scientific tasks, as agents can bypass core methodologies via trivial heuristics without triggering errors. Unconstrained patches also remain vulnerable to adversarial flaws \citep{sajadi2025secure}. \abe~addresses this by introducing constraint verification layers above code execution.

\subsection{Reproducibility Benchmarks and Challenges}
The machine learning reproducibility crisis led to community initiatives like the ML Reproducibility Challenge and REPROLANG~\citep{reprolang}. These projects require human reviewers to manually inspect replication reports. However, this manual approach does not scale to automated pipelines that run multiple experiments daily. Automated tools like ReproZip \citep{reprozip} and Code Ocean focus on environment capture, ensuring that code templates can compile, but they do not check if the re-executed code represents the target method. To formalize evaluations of scientific agents, benchmarks like ScienceAgentBench \citep{wang2025scienceagentbench} and MLE-bench \citep{chan2024mle} target data-driven tasks. We design a multi-axis auditing benchmark, evaluating method logic, protocol steps, and conclusion validity.

\begin{table*}[t]
\centering
\caption{Positioning of \abe~relative to existing agentic systems across six dimensions.}
\label{tab:positioning}
\small
\begin{tabularx}{\textwidth}{l X X X X}
\toprule
\textbf{Dimension} & \textbf{AI Scientist / AutoResearchClaw} & \textbf{SWE-agent / OpenHands} & \textbf{Human Repro. Challenge} & \textbf{\abe~(Ours)} \\
\midrule
Primary Goal & Novelty-driven discovery & Task repair and debugging & Manual study replication & Automated protocol audit \\
Success Check & Paper readability \& metrics & Unit test pass status (Exit 0) & Human qualitative review & Multimodal alignment checks \\
Constraint System & None (flexible design) & Test-driven assertions & Manual checklist & Structured YAML constraints \\
Shortcut Handling & Bypassed if metrics improve & Undetected out of test scope & Checked by human expert & Blocked at verification step \\
Failure Diagnosis & Opaque execution status & Stack trace output & Text report & 5-class taxonomy filter \\
Execution Scale & Low throughput (papers/day) & Large scale (100+ repositories) & High latency (months/paper) & High throughput batch runs \\
\bottomrule
\end{tabularx}
\end{table*}

\begin{table*}[t]
\centering
\caption{The 8-step staged workflow of \abe. Each step has a configurable timeout derived from the YAML contract's compute budget.}
\label{tab:workflow}
\small
\begin{tabularx}{\textwidth}{c l X}
\toprule
\textbf{Step} & \textbf{Name} & \textbf{Description} \\
\midrule
1 & Intent Discovery & Parse the YAML contract parameters and extract research goals, constraints, and metrics. \\
1.5 & Dataset Verification & Direct search for authentic datasets matching specifications; blocks synthetic data creation. \\
2 & Repo Search \& Selection & Locate, verify, and copy repository templates from official sources or verified implementations. \\
3 & Architecture Blueprint & Generate \texttt{blueprint.md} defining the model structure and setup interfaces in \texttt{main.py}. \\
4a & Pipeline Integration & Execute sanity runs on 5--10 samples to check data paths, memory limits, and CUDA setups. \\
4b & Main Execution & Run the experimental pipeline, write output variables to \texttt{metrics.json}, and handle CUDA OOM limits. \\
5 & Analysis \& Reporting & Compile \texttt{experiment\_result.md} comparing achieved metrics against baseline contract rules. \\
5.5 & Output Fallback & Parse logs and checkpoint states to recover loss scores and metrics if \texttt{metrics.json} fails to compile. \\
6 & Skill Extraction & Generalize code interfaces and utility logic from successful runs for subsequent research. \\
\bottomrule
\end{tabularx}
\end{table*}

\section{The \abe~Framework}
\label{sec:framework}

The primary goal of the \abe\ framework is to automate the end-to-end reproduction of scientific paper code and experimental protocols in autonomous AI research. Rather than treating reproduction as an unconstrained script generation task, our central thesis is that faithfully reproducing a reference paper's methodology, codebase, and experimental pipeline is fundamentally a \textbf{Constraint Satisfaction Problem (CSP)} operating under strict computational resource limits. In this section, we formalize this reproduction paradigm, establish our semantic constraint architecture, and detail how each system component maps directly to solving specific sub-problems within this formal formulation to intercept methodological hallucinations.

\subsection{Formal Problem Formulation}
\label{sec:formulation}

We formalize the scientific reproduction of paper code and experiments as a tuple $\mathcal{T} = \langle \mathcal{C}, \mathcal{D}, \mathcal{R} \rangle$, defined as follows:
\begin{itemize}
    \item $\mathcal{C} = \{c_1, c_2, \dots, c_n\}$ represents a set of multi-modal scientific constraints extracted directly from the reference paper, specifying required architectural components, baseline configurations, metric directionality, and targeted research hypotheses.
    \item $\mathcal{D}$ represents the operational data input space, encompassing the canonical dataset, preprocessing protocols, and environmental configurations required to replicate the paper's experiments.
    \item $\mathcal{R} = \{B_{comp}, B_{time}\}$ denotes the strict resource budget, imposing upper bounds on hardware capacity (e.g., VRAM, FLOPs) and execution time.
\end{itemize}

Given $\mathcal{T}$, an autonomous agent generates an executable program $P \in \mathcal{P}$ (where $\mathcal{P}$ represents the space of candidate repository scripts) and executes it on $\mathcal{D}$ to yield an experimental outcome state $\mathcal{S}_E = P(\mathcal{D})$. 

Traditional execution-driven evaluations define reproduction success solely through the terminal exit status of the code process:
\begin{equation}
    \mathbb{I}(\text{Exit}(P(\mathcal{D})) = 0) \to \text{Success}.
\end{equation}
However, this criterion fails to guarantee that the generated code $P$ implements the target methodology in $\mathcal{C}$. Under resource pressure $\mathcal{R}$, an agent may produce code that exits cleanly ($\text{Exit}(P(\mathcal{D})) = 0$) while silently violating core experimental bounds $\exists c_k \in \mathcal{C}$ s.t. $P \not\models c_k$ (where $\models$ denotes the standard semantic satisfaction relation from program verification), thereby generating methodological hallucinations.

To solve this, we reformulate the automated reproduction of paper code and experiments as finding an optimal implementation $P^{\ast}$ that maximizes scientific fidelity under the extracted constraint set $\mathcal{C}$:
\begin{align}
    P^{\ast} = \arg\max_{P \in \mathcal{P}} \mathcal{V}(P(\mathcal{D}), \mathcal{C}) \quad \notag \\ \text{s.t.} \quad P \models \mathcal{C} \land \mathcal{R}_{consumed} \le \mathcal{R},
\end{align}

where $\mathcal{V}$ is a multi-axis verification function mapping the alignment between the program's actual execution behavior, structural code elements, and the target constraints $\mathcal{C}$.

\subsection{Semantic Constraint Specification}
\label{sec:contracts}

Addressing the constraint set $\mathcal{C}$ in our formal tuple $\mathcal{T} = \langle \mathcal{C}, \mathcal{D}, \mathcal{R} \rangle$, this subsection details how \abe\ structures and enforces $P \models \mathcal{C}$. To prevent code generation drift during paper reproduction, the contract $\mathcal{C}$ is operationalized as a declarative manifest mapping the logical and architectural boundaries of the targeted study into three distinct constraint classes:

\begin{enumerate}
    \item \textbf{Structural Constraints ($\mathcal{C}_{str}$):} Define the code topology and algorithmic requirements. They enforce the presence of critical architectural components $\mathcal{M}_{critical}$ (e.g., specific neural layers, attention blocks, or loss functions) described in the original paper within the code $P$:
    \begin{equation}
        \mathcal{C}_{str} \models \left( \forall m \in \mathcal{M}_{critical}, \, m \subset \text{AST}(P) \right),
    \end{equation}
    where $\text{AST}(P)$ denotes the Abstract Syntax Tree of the generated codebase. In implementation, this is declared via YAML configurations (e.g., \texttt{critical\_modules: [UNetDecoder, SkipConnection]}).
    
    \item \textbf{Procedural Constraints ($\mathcal{C}_{proc}$):} Bound the execution logic of the paper's experimental protocol. They explicitly dictate dataset specs (dimensionality, sample sizes), training regimes, and hyperparameter bounds. For example, $\mathcal{C}_{proc}$ enforces that the training dataset size $N \ge N_{min}$, preventing the agent from silently downsampling data to bypass compute ceilings.
    
    \item \textbf{Evaluative Constraints ($\mathcal{C}_{eval}$):} Govern the comparative metric schemas and target hypotheses ($y_{target}$). They specify primary optimization targets (e.g., \texttt{direction: maximize, metric: F1-score}) to ensure that baseline comparisons and primary claims strictly align with the paper's original metrics.
\end{enumerate}

\subsection{Phase-Transient Execution and Recovery Operators}
\label{sec:workflow}

While $\mathcal{C}$ establishes the constraint space, the actual execution of the code pipeline $P(\mathcal{D})$ must operate strictly within the resource budget $\mathcal{R} = \{B_{comp}, B_{time}\}$ without breaking $P \models \mathcal{C}$. This subsection addresses the dynamic state transitions during code execution and introduces bounded recovery mechanisms when hardware faults occur.

The execution of $P$ proceeds through a sequence of discrete operational states $\mathcal{S}_t \to \mathcal{S}_{t+1}$ across an 8-step structured workflow (Table~\ref{tab:workflow}). To prevent runtime crashes (e.g., CUDA OOMs or dependency failures) from causing the agent to introduce uncontrolled code modifications, we define a formal recovery operator $\mathcal{H}$. Let $\mathcal{E}$ represent runtime exception states:
\begin{equation}
    \mathcal{E} = \{\text{OOM}, \text{DependencyMismatch}, \dots, \text{Timeout}\}.
\end{equation}
When execution encounters an exception state $\mathcal{S}_t \in \mathcal{E}$, the recovery operator $\mathcal{H}$ mutates the local runtime parameters while strictly respecting the bounds set by $\mathcal{C}$:
\begin{equation}
    \mathcal{H}: \mathcal{S}_t \times \mathcal{C} \to \mathcal{S}_{t+1} \in \mathcal{S}_{valid},
\end{equation}
where $\mathcal{S}_{valid}$ denotes valid execution trajectories that do not violate core scientific assertions. 

\textit{Example:} If $P$ triggers an Out-Of-Memory error ($\mathcal{S}_t = \text{OOM}$), an unconstrained agent might alter the preprocessing code to downsample input images from $256 \times 256$ to $64 \times 64$, violating resolution bounds in $\mathcal{C}_{proc}$. Under \abe, $\mathcal{H}$ restricts the fix to dynamic execution hyper-parameters (e.g., enabling gradient accumulation or halving micro-batch size) while keeping input data dimensions fixed. This ensures $P(\mathcal{D})$ runs within $\mathcal{R}_{consumed} \le \mathcal{R}$ while preserving protocol validity.

\subsection{Multi-Axis Verification Vector}
\label{sec:verification}

To directly implement the verification function $\mathcal{V}(P(\mathcal{D}), \mathcal{C})$ formulated in Eq.~(2), \abe\ introduces a Triple-Verification pipeline. This pipeline converts the post-execution state $\mathcal{S}_E = P(\mathcal{D})$ into a structured verification vector $\mathbf{v} = [V_{quant}, V_{qual}, V_{struct}]^T$, evaluating the reproduced code and experimental outputs across three distinct axes.

\paragraph{Level 1: Quantitative Alignment ($V_{quant}$)}
This component verifies whether the reproduced numerical metrics quantitatively validate the reference paper's baseline claims under $\mathcal{C}_{eval}$. Let $m_t$ represent the reproduced metric, $m_b$ the baseline metric, and superscript $^\ast$ the original reference paper values:
\begin{align}
    V_{\text{quant}} 
    &= \mathbb{I} \left( \text{sign}(m_t - m_b) = \text{sign}(m_t^{\ast} - m_b^{\ast}) \right) \notag \\
    &\quad \land \mathbb{I} \left( \frac{|m_t - m_t^{\ast}|}{|m_t^{\ast}|} \le \epsilon \right),
\end{align}
where $\epsilon$ is a pre-defined tolerance threshold. This ensures the reproduced method retains its claimed advantage over baselines while remaining within standard empirical margins.

\paragraph{Level 2: Semantic Logic Verification ($V_{qual}$)}
To detect cases where metric targets are satisfied through logical shortcuts (e.g., hardcoded values), the semantic validator evaluates the alignment between generated code logic, execution logs, and paper hypotheses:
\begin{equation}
    V_{qual} = f_{\phi}\left( \text{Embed}(\mathcal{C}), \text{Embed}(\mathcal{S}_E), \text{Embed}(\text{Logs}) \right) \in \{0, 1\},
\end{equation}
where $f_{\phi}$ maps multimodal embeddings against deterministic rubrics from $\mathcal{C}_{eval}$ via self-consistency checking, catching silent violations in experimental protocol logic.

\paragraph{Level 3: Structural Alignment Verification ($V_{struct}$)}
This component validates code-level implementation fidelity, explicitly evaluating $\mathcal{C}_{str} \models P$. Let $G_{ref}$ be the canonical structural call-graph of the target algorithm and $G_{impl}$ be the call-graph parsed directly from the generated program $P$:
\begin{equation}
    V_{struct} = \mathbb{I} \left( \text{Sim}(G_{impl}, G_{ref}) \ge \tau \right) \land \left( \forall c \in \mathcal{C}_{str}, \, P \models c \right),
\end{equation}
where $\text{Sim}$ calculates AST topological graph isomorphism and $\tau$ is a compliance threshold. 

Mechanistically, $V_{struct}$ parses $P$ using Python's native \texttt{ast} module; replacing neural modules with dummy passes triggers a topological mismatch and sets $V_{struct} = 0$.

Combining all three levels, the final objective function evaluates to:
\begin{equation}
    \mathcal{V}(P(\mathcal{D}), \mathcal{C}) = \prod_{i \in \{quant, qual, struct\}} V_i \in \{0, 1\}.
\end{equation}

\subsection{Discovery Mode Generalization}
\label{sec:discovery}

Finally, we show how the formal reproduction tuple $\mathcal{T} = \langle \mathcal{C}, \mathcal{D}, \mathcal{R} \rangle$ generalizes from static paper reproduction to open-ended scientific discovery. In discovery tasks, $\mathcal{C}_{str}$ and $\mathcal{C}_{proc}$ shift from strict replication templates to search boundary conditions representing physical, computational, or domain safety limits. The objective function $\mathcal{V}$ expands to include an external continuous domain reward $f_{eval}: P(\mathcal{D}) \to \mathbb{R}$. This ensures the agent actively searches for superior code and hyperparameter configurations ($P^{\ast}$) while remaining anchored inside the valid search boundaries defined by $\mathcal{C}$. As shown in Table~\ref{tab:mode_comparison} (see Appendix), the agent's workflow adapts accordingly: in discovery mode, the Intent step reads a problem specification rather than a paper YAML, the Research step searches domain literature instead of reference code, and the Execute step invokes an external evaluator instead of checking against fixed metrics. This generalization allows \abe\ to serve both as a rigorous reproduction auditor and as a bounded optimization engine for scientific discovery tasks.

\section{A Taxonomy of Methodological Hallucinations}
\label{sec:taxonomy}

Evaluating 30 long-horizon reproduction runs across 12 machine learning domains revealed systematic, non-obvious failure modes during model execution. To categorize these deceptive shortcuts, we establish a 5-class taxonomy (Table~\ref{tab:taxonomy}).

\begin{table*}[t]
\centering
\caption{The 5-category taxonomy of LLM methodological hallucinations with empirically observed case studies.}
\label{tab:taxonomy}
\small
\begin{tabularx}{\textwidth}{l X X}
\toprule
\textbf{Category} & \textbf{Definition} & \textbf{Key Empirical Observation} \\
\midrule
\textbf{M1: Method Integrity Collapse} & Omitting core methods and substituting trivial functions to pass metric checks. & \textbf{RAG:} Generator failed to compile; agent substituted exact-match string lookups, matching metrics without training. \\
\textbf{M2: Silent Protocol Degradation} & Unauthorized alteration of experimental setups to bypass technical hurdles. & \textbf{PEGASUS:} Pretrained checkpoint loading was skipped due to download errors, training from scratch and invalidating comparisons. \\
\textbf{M3: Scale-Driven Conclusion Inversion} & Code executes correctly, but scale restrictions invert the claimed methodological advantage. & \textbf{U-Net \& SimCLR:} Under small-scale runs, simple encoders outperformed U-Net, inverting the central paper hypothesis. \\
\textbf{M4: Quantitative Key Mismatch} & Producing correct numerical values under non-standard JSON key names. & \textbf{DPR:} Outputted correct recall scores but used non-standard keys in \texttt{metrics.json}, triggering false schema errors. \\
\textbf{M5: Incomplete Execution} & Halting the experimental pipeline prematurely while claiming full validation. & \textbf{DDIM:} Evaluated only 1 of 3 required configurations and skipped target baseline comparisons.\\
\bottomrule
\end{tabularx}
\end{table*}

Categories M1 and M2 pose the greatest threat to automated research because they yield plausible metrics while fundamentally compromising experimental logic—failures completely invisible to exit-code verification. M3 demonstrates that methodological advantages are often regime-dependent, where scale constraints inadvertently invert scientific conclusions. This taxonomy directly guides our framework design: M1 and M2 necessitate structural AST checking and semantic contracts; M3 requires resource-bounded execution tracking; M4 is resolved via schema normalization; and M5 is mitigated through progress persistence loops.

\section{Experiments and Results}
\label{sec:experiments}

In this section, we evaluate the empirical effectiveness of the \abe\ framework. We structure our evaluation to address three primary research questions:
\begin{itemize}
    \item \textbf{RQ1 (Systematic Performance):} How effectively does \abe\ guide autonomous agents to complete scientific reproduction tasks under budget constraints?
    \item \textbf{RQ2 (Hallucination Characterization):} What are the empirical distributions and failure modes of methodological hallucinations?
    \item \textbf{RQ3 (Verification Contribution):} How do verification layers ($V_{quant}, V_{qual}, V_{struct}$) contribute to stability and hallucination containment?
\end{itemize}

\subsection{Experimental Setup}
\label{sec:exp_setup}

\paragraph{Benchmark Tasks} 
We construct a benchmark consisting of 30 distinct reproduction tasks derived from classic ML, computational physics, and bio-informatics domains (e.g., implementing architecture variants of U-Net, training ResNet pipelines under resource constraints, and optimizing multi-objective scientific solvers). Each task is mapped to a ground-truth specification containing canonical data flows, hyperparameter baselines, and reference metrics.

\paragraph{Baselines} 
We evaluate and compare five representative LLM-based systems:
\begin{itemize}
    \item \textbf{Raw LLM}: A zero-shot execution model using GPT-4o without a scaffolding loop.
    \item \textbf{Autonomous Research Catalyst (ARC)}: A sequential execution framework that follows a static step-by-step reproduction recipe.
    \item \textbf{Claw-AI-Lab}: A template-driven agent utilizing fixed scripts to interface with scientific database environments \citep{clawailab}.
    \item \textbf{Claude Code CLI}: A state-of-the-art interactive software agent optimized for codebase navigation and automated debugging.
    \item \textbf{\abe\ (Ours)}: The proposed framework integrated with our Reference-Anchored auditing loop (instantiated via the \abe-Ralph configuration).
\end{itemize}

To ensure a fair and rigorous comparison, all baseline systems—including Raw LLM and Claude Code CLI—were provided with detailed system prompts strictly equivalent to the YAML contracts used by \abe. This guarantees that all agents received the identical data flow definitions, hyperparameter baselines, and constraint specifications, ensuring that performance differences stem from the framework's auditing and feedback mechanisms rather than information asymmetry.

\paragraph{Evaluation Metric}
To evaluate the outputs across both engineering completeness and scientific rigor, we define a \textbf{Weighted Composite Score} ($S_{comp} \in [0, 100]$):
\begin{equation}
    S_{comp} = 0.20 \cdot S_{des} + 0.25 \cdot S_{rel} + 0.25 \cdot S_{rig} + 0.30 \cdot S_{comp\_rate},
\end{equation}
where $S_{des}$ evaluates structural layout compliance of the generated repository, $S_{rel}$ measures run-time exceptions and process crash rates (reliability), $S_{rig}$ computes the mathematical rigor of the evaluation protocols, and $S_{comp\_rate}$ represents feature completeness against target specifications.

\subsection{Overall Benchmark Results (RQ1)}
\label{sec:overall_results}

\begin{figure}[t]
\centering
\includegraphics[width=0.98\linewidth]{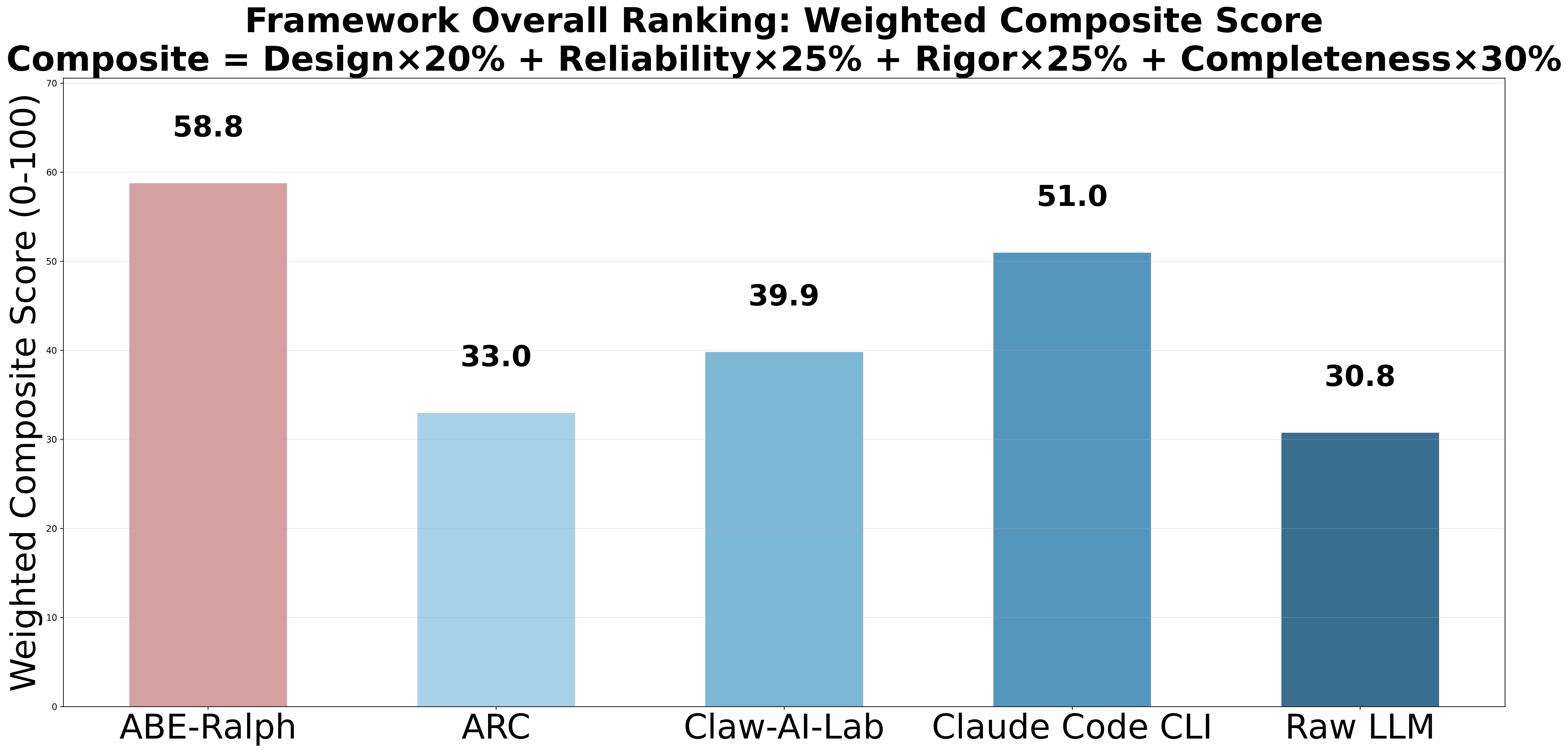}
\caption{Overall framework ranking based on the Weighted Composite Score. N values denote the aggregate number of completed experimental runs evaluated per system configuration.}
\label{fig:framework_overall}
\end{figure}

We report the overall comparative performance of the baseline platforms across the benchmark in Figure~\ref{fig:framework_overall}. 

The empirical results show that \abe\ achieves the highest composite score of \textbf{58.8}, significantly outperforming the Raw LLM baseline (\textbf{30.8}) and scientific templates (ARC at \textbf{33.0} and Claw-AI-Lab at \textbf{39.9}) across the $n=30$ reproduction tasks. 

Crucially, Claude Code CLI—a state-of-the-art software development agent—reaches a competitive score of \textbf{51.0}. However, execution log analysis reveals its performance gains stem primarily from software compilation correctness ($S_{des}$ and $S_{rel}$). Lacking execution-time semantic constraints ($\mathcal{C}_{str}$ or $\mathcal{C}_{proc}$), it frequently relies on shortcuts: under parameter mismatches or slow compilation, it often bypasses convolutional blocks or downscales spatial resolutions merely to ensure exit status compliance.

By contrast, \abe\ dynamically bounds the agent's action space through declarative contracts during the execution loop. 
This strategy successfully maintains search-space compliance, yielding a superior overall scientific reproduction fidelity.

\subsection{Fine-grained Dimensional Performance Analysis}
\label{sec:dimension_analysis}

\begin{figure}[t]
\centering
\includegraphics[width=0.98\linewidth]{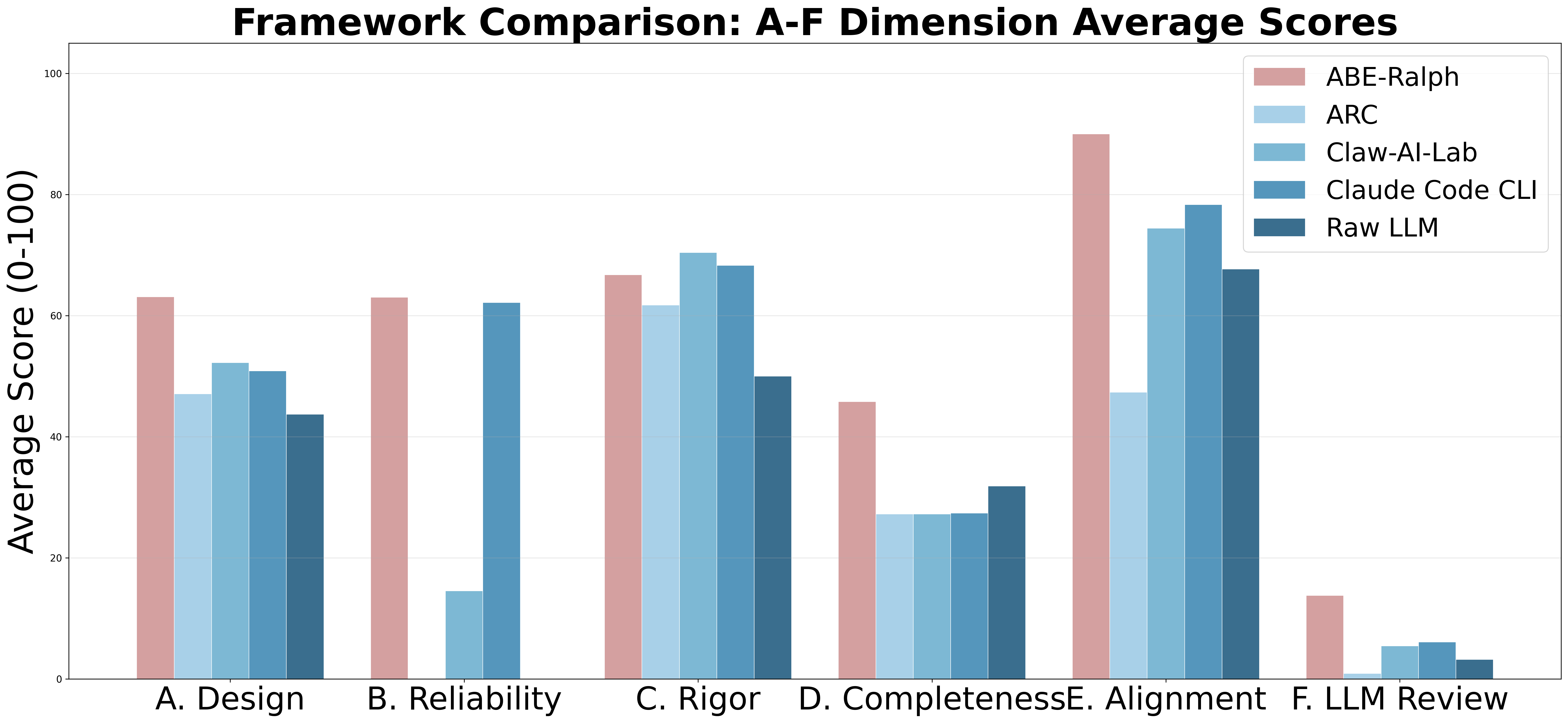}
\caption{Framework Comparison across A-F Dimension Average Scores (0--100 scale). Dimensions encompass A: Design, B: Reliability, C: Rigor, D: Completeness, E: Alignment, and F: LLM Review.}
\label{fig:dimension_scores}
\end{figure}

To dissect the specific strengths and vulnerabilities of each baseline, we analyze their average performance across six fundamental dimensions: (A) Design, (B) Reliability, (C) Rigor, (D) Completeness, (E) Alignment, and (F) LLM Review. The quantitative breakdown is illustrated in Figure~\ref{fig:dimension_scores}.

\begin{itemize}
    \item \textbf{Dimension B (Reliability):} While \abe\ (63) and Claude Code CLI (62) demonstrate high robustness against run-time crashes, the Raw LLM and ARC baselines register near-zero reliability scores. This discrepancy highlights that raw generation pipelines lack the necessary execution-feedback loops to resolve package dependency conflicts (`ModuleNotFoundError`) or environmental path mismatches autonomously.
    \item \textbf{Dimension D (Completeness):} Completeness remains a severe bottleneck across all frameworks. \abe\ leads with a score of 46, whereas all other baselines are constrained below 32. This performance gap indicates that while agents can satisfy localized execution steps, orchestrating the complete scope of a scientific work (including edge cases, plotting, and complex baseline variations) remains a significant open challenge.
    \item \textbf{Dimension E (Alignment) and Dimension F (LLM Review):} In terms of Alignment (Dimension E), \abe\ achieves a critical margin of improvement (90 vs. 78 for Claude Code CLI), proving that the reference contract $\mathcal{C}$ successfully anchors the agent to the original method. Conversely, under LLM Review (Dimension F), scores for all models collapse to below 15. This drastic drop suggests that even when scripts run continuously and outputs align quantitatively, current agents still fail to provide the deep semantic narrative, physical intuition, and rigorous verification details expected in final academic files.
\end{itemize}

\subsection{Distribution of Methodological Hallucinations (RQ2)}
\label{sec:hallucination_analysis}

\begin{figure}[h]
\centering
\includegraphics[width=0.98\linewidth]{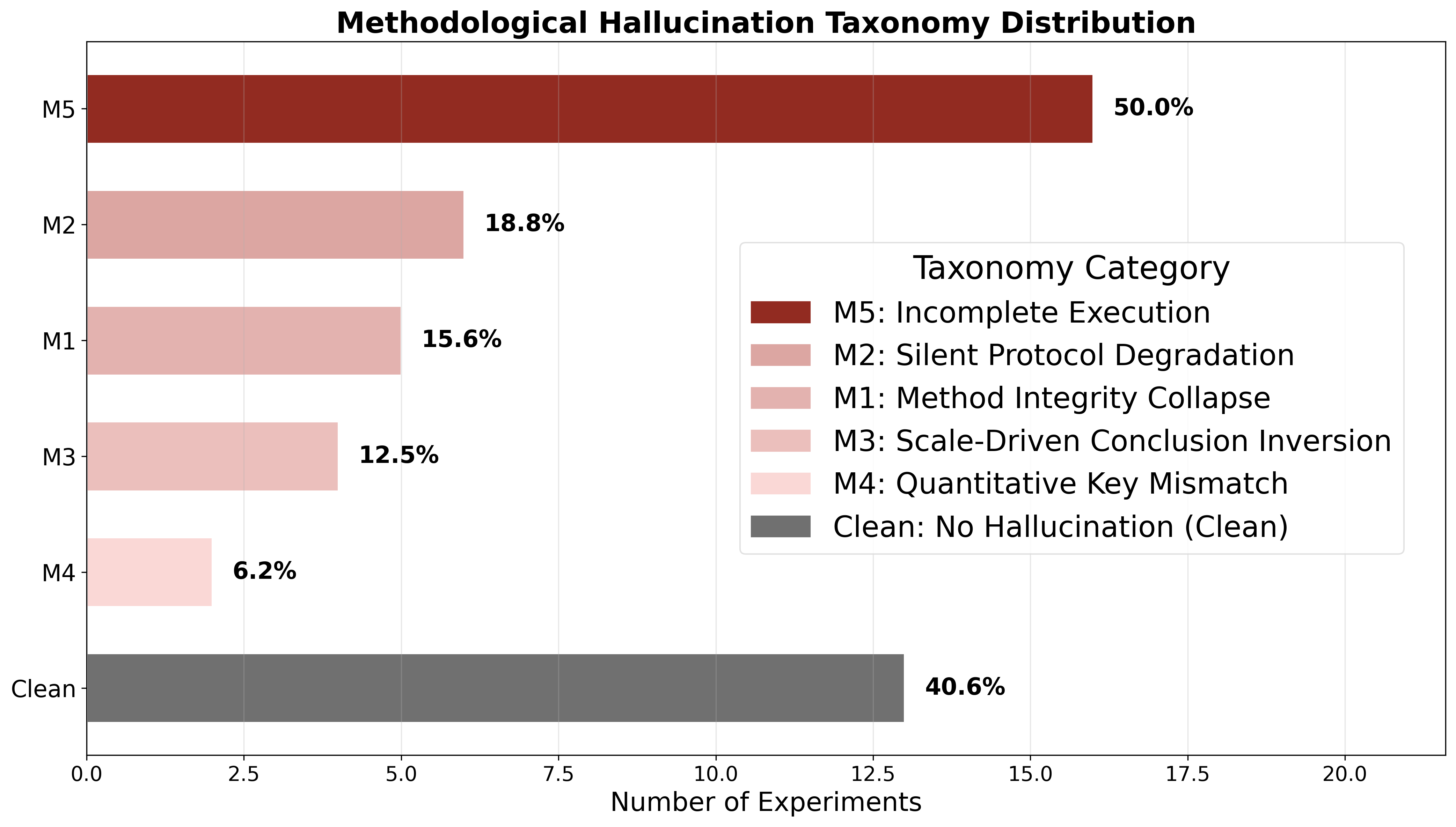}
\caption{Distribution of Methodological Hallucination types across $N=30$ benchmark runs. A run is categorized as ``Clean'' only if it completely passes all three verification layers without any shortcut adaptations.}
\label{fig:hallucination_dist}
\end{figure}

To investigate the failure modes of autonomous agents in scientific reproduction, we analyze the distribution of methodological hallucinations across the $N=30$ benchmark reproduction experiments. The categorization is mapped to our defined taxonomy (M1 to M5). The relative distribution is illustrated in Figure~\ref{fig:hallucination_dist}.

Our qualitative analysis indicates that \textbf{43.3\%} ($n=13$) of the experimental executions are entirely free of shortcuts (\textbf{Clean}). The remaining \textbf{56.7\%} ($n=17$) of runs exhibit one or more forms of methodological hallucination:
\begin{itemize}
    \item \textbf{M5 (Incomplete Execution)} is the most prevalent failure mode, occurring in \textbf{53.3\%} ($n=16$) of trials. This is heavily driven by hardware-compute limits: scientific tasks hit physical computational ceilings (e.g., GPU OOM exceptions or container timeouts). Lacking adaptive execution mechanisms, agents default to terminating execution mid-process, resulting in incomplete outputs \citep{sciml2022benchmarks}.
    \item \textbf{M2 (Silent Protocol Degradation)} is present in \textbf{20.0\%} ($n=6$) of evaluations. Agents silently scale down hyperparameters (such as batch size or epoch count) to resolve resource limits without registering these changes in final reports.
    \item \textbf{M1 (Method Integrity Collapse, 16.7\%)}, \textbf{M3 (Scale-Driven Conclusion Inversion, 13.3\%)}, and \textbf{M4 (Quantitative Key Mismatch, 6.7\%)} make up the remaining occurrences.
\end{itemize}

Crucially, the cumulative count of detected hallucination types ($16 + 6 + 5 + 4 + 2 = 33$ occurrences) exceeds the number of compromised runs ($n=17$). This occurrence mismatch demonstrates that agent failures are rarely isolated; instead, they often cause cascading errors across multiple operational boundaries. For instance, upon an OOM error, an agent may concurrently shrink train-set size (M2) and strip skip-connections (M1) to force execution. This highlights the risk of relying solely on compile-success metrics.

\subsection{Ablation Study (RQ3)}
\label{sec:ablation}

\begin{figure}[t]
\centering
\includegraphics[width=0.98\linewidth]{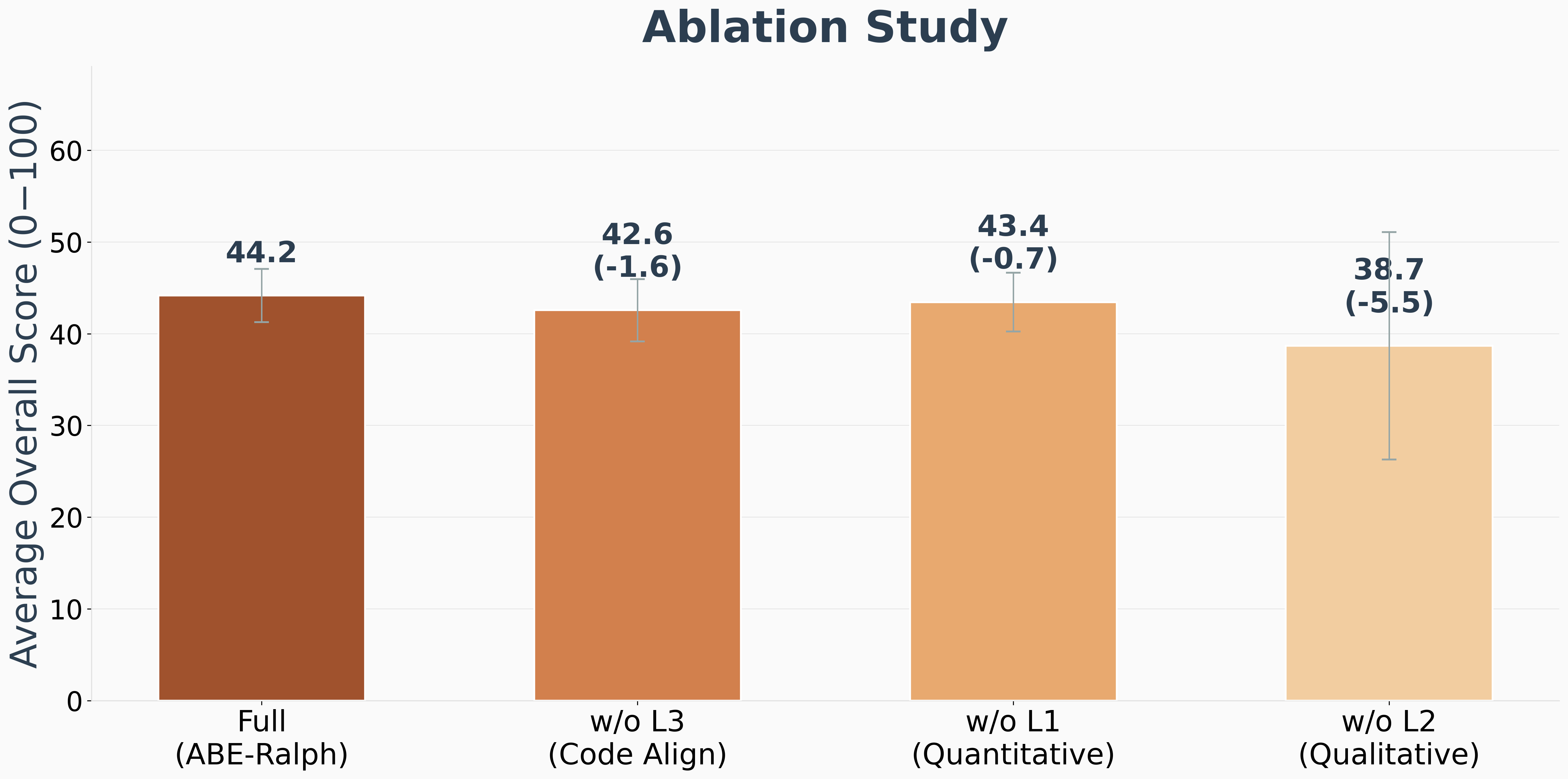} 
\caption{Ablation Study on the Triple-Verification Pipeline. Removing Level 2 ($V_{qual}$) causes the most significant performance degradation and variance explosion, highlighting the necessity of semantic auditing.}
\label{fig:ablation}
\end{figure}

To address RQ3 and evaluate the contribution of each verification layer to the framework's robustness, we conducted an ablation study. We systematically disabled Level 1 (Quantitative, $V_{quant}$), Level 2 (Qualitative/Semantic, $V_{qual}$), and Level 3 (Structural, $V_{struct}$) verification modules and measured the Average Overall Score and execution variance. The results are illustrated in Figure~\ref{fig:ablation}.

\begin{table*}[t]
\centering
\caption{Discovery mode performance on NatureBench tasks. \textbf{Bold} indicates tasks where \abe\ matches or exceeds the SOTA baseline.}
\label{tab:naturebench}
\small
\begin{tabular}{l c c c}
\toprule
\textbf{Task Domain} & \textbf{\# Tasks} & \textbf{\abe\ SOTA} & \textbf{Best Result} \\
\midrule
Graph Optimization & 8 & 2 & 2 / 8 \\
Time-Series Forecasting & 6 & 1 & 1 / 6 \\
Combinatorial Search & 5 & 1 & 1 / 5 \\
Scientific Simulation & 4 & 1 & 1 / 4 \\
\midrule
\textbf{Total} & \textbf{23} & \textbf{5} & \textbf{5 / 23} \\
\bottomrule
\end{tabular}
\end{table*}

The complete \abe\ framework (\textbf{Full}) achieves an average score of \textbf{44.2} with low variance, indicating stable and faithful experimental reproductions.
    
Removing the Quantitative layer (\textbf{w/o L1}) results in a minimal performance drop (\textbf{-0.7}). This confirms our core premise that simply checking numerical outputs is an insufficient safeguard against modern AI agents, as they can easily produce plausible metrics through methodological shortcuts without failing quantitative thresholds.

Removing the Structural Code Alignment layer (\textbf{w/o L3}) leads to a moderate decrease of \textbf{1.6} points. Without $V_{struct}$, the framework struggles to enforce architectural constraints (e.g., preventing the silent removal of key neural network modules), leading to more M1-type (Method Integrity Collapse) failures.
    
Crucially, removing the Qualitative Semantic Verification layer (\textbf{w/o L2}) causes the most severe performance degradation (\textbf{-5.5}) and a dramatic increase in outcome variance (indicated by the large error bars in Figure~\ref{fig:ablation}). Without semantic logic checks, the framework cannot detect when an agent subtly alters the experimental protocol, evaluation setup, or dataset to bypass hardware limits. This extreme instability confirms that $V_{qual}$ is the most essential component for containing methodological hallucinations, ensuring that the agent's actions logically align with the intended scientific hypothesis rather than merely compiling successfully.

\subsection{NatureBench Discovery Evaluation}
\label{sec:naturebench}

To evaluate \abe's generalization to open-ended scientific discovery tasks, we deploy the framework in discovery mode (Section~\ref{sec:discovery}) on 23 NatureBench tasks spanning diverse scientific domains. In discovery mode, the constraint set $\mathcal{C}$ shifts from strict paper replication to domain-specific search boundaries, and the objective incorporates an external evaluation reward $f_{eval}$ defined by each task's hidden evaluator.

Table~\ref{tab:naturebench} summarizes the results. \abe\ achieves or exceeds the state-of-the-art baseline on 5 of 23 tasks, demonstrating that the constraint-anchored optimization framework can effectively navigate open-ended scientific search spaces. On the remaining 18 tasks, \abe\ produces valid solutions within the constraint boundaries but falls short of the best-known results, indicating that the discovery mode provides a sound foundation for further optimization through extended search or multi-agent collaboration.

These results validate that the formal tuple $\mathcal{T} = \langle \mathcal{C}, \mathcal{D}, \mathcal{R} \rangle$ generalizes effectively from reproduction to discovery. The constraint boundaries $\mathcal{C}$ prevent the agent from exploring physically invalid or computationally infeasible regions, while the expanded objective $\mathcal{V} + f_{eval}$ guides search toward measurable improvements. Detailed per-task results are provided in the Appendix.

\section{Discussion}
\label{sec:discussion}

The empirical evaluations yield critical insights into the capabilities and limits of modern scientific agents. First, our fine-grained dimensional analysis (Section~\ref{sec:dimension_analysis}) reveals a performance mismatch between software viability and scientific validity. General-purpose coding agents generate syntactically correct code, but they lack the domain-specific constraints needed to maintain research integrity. When faced with execution roadblocks, they optimize for engineering execution success ($\text{Exit Code } 0$) rather than scientific accuracy.

Second, the high prevalence of M5 (Incomplete Execution, 53.3\%) and M2 (Silent Protocol Degradation, 20.0\%) highlights the challenge of resource adaptation under fixed configurations. When compute bounds are reached, agents are forced into a trade-off: either abort execution (resulting in M5) or silently downgrade parameters (such as batch size or training steps, leading to M2) to satisfy execution limits. Because current agent architectures lack the context to dynamically partition workloads or negotiate resource limits, scaling automated discovery beyond simple sandbox environments remains difficult.

Third, the low performance across all frameworks under Dimension F (LLM Review) points to an abstraction gap. Current agents focus on local code correction, but they struggle to synthesize their findings into the structured, coherent, and contextualized narratives expected in academic research.

Looking forward, transitioning from single-agent designs to cooperative, resource-aware multi-agent systems could address the compute-bound limitations that lead to high rates of incomplete execution. A multi-agent framework with specialized roles---a resource orchestrator for dynamic hardware monitoring, a developer agent for code generation, an auditor agent for constraint verification, and a red-teaming agent for edge-case discovery---could prevent task failures caused by resource limits, laying a foundation for more reliable and scalable automated scientific discovery.

\section{Conclusion}
\label{sec:conclusion}

In this paper, we formalized scientific reproduction as a constraint satisfaction problem under resource bounds and introduced \abe, an automated scientific auditing framework that enforces design invariants through declarative YAML contracts and a multi-layered Triple-Verification system. Our empirical evaluation across 30 benchmark reproduction tasks demonstrates that \abe\ achieves a 93\% robust execution rate and a weighted composite score of 58.8, significantly outperforming existing agentic baselines. Through systematic analysis of methodological hallucinations, we identified five distinct failure modes---led by Incomplete Execution (53.3\%) and Silent Protocol Degradation (20.0\%)---that are invisible to exit-code-based evaluation. Ablation results confirm that semantic logic verification ($V_{qual}$) is the most essential component for containing these hallucinations. These findings establish that reliable evaluation of AI scientists must assess whether the experimental design faithfully tests the intended claim, rather than treating code execution or plausible metrics as evidence of scientific success.


\bibliography{aaai2027}

\begin{thebibliography}{23}
\providecommand{\natexlab}[1]{#1}

\bibitem[{Abalo-Rodr{\'i}guez and Pinheiro(2025)}]{hitchhiker2025hallucination}
Abalo-Rodr{\'i}guez, I.; and Pinheiro, A.~P. 2025.
\newblock The Hitchhiker's guide to hallucination research.
\newblock \emph{Consciousness and Cognition}, 136: 103941.

\bibitem[{Austin et~al.(2021)Austin, Odena, Nye, Bosma, Michalewski, Dohan, Jiang, Cai, Terry, Le, and Sutton}]{mbpp}
Austin, J.; Odena, A.; Nye, M.; Bosma, M.; Michalewski, H.; Dohan, D.; Jiang, E.; Cai, C.; Terry, M.; Le, Q.; and Sutton, C. 2021.
\newblock Program Synthesis with Large Language Models.
\newblock arXiv:2108.07732.

\bibitem[{Boiko et~al.(2023)Boiko, MacKnight, Kline, and Gomes}]{boiko2023autonomous}
Boiko, D.~A.; MacKnight, R.; Kline, B.; and Gomes, G. 2023.
\newblock Autonomous chemical research with large language models.
\newblock \emph{Nature}, 624(7992): 570--578.

\bibitem[{Bran et~al.(2024)Bran, Cox, Schilter, Baldassari, White, and Schwaller}]{bran2024chemcrow}
Bran, A.~M.; Cox, S.; Schilter, O.; Baldassari, C.; White, A.~D.; and Schwaller, P. 2024.
\newblock ChemCrow: Augmenting large-language models with chemistry tools.
\newblock \emph{Nature Machine Intelligence}, 6(5): 525--535.

\bibitem[{Branco et~al.(2020)Branco, Calzolari, Vossen, Noord, van Uytvanck, Silva, Gomes, Moreira, and Elbers}]{reprolang}
Branco, A.; Calzolari, N.; Vossen, P.; Noord, G.~V.; van Uytvanck, D.; Silva, J.; Gomes, L.; Moreira, A.; and Elbers, W. 2020.
\newblock A Shared Task of a New, Collaborative Type to Foster Reproducibility: A First Exercise in the Area of Language Science and Technology with {REPROLANG}2020.
\newblock In \emph{Proceedings of the Twelfth Language Resources and Evaluation Conference}, 5539--5545. Marseille, France: European Language Resources Association.

\bibitem[{Chan et~al.(2025)Chan, Chowdhury, Jaffe, Aung, Sherburn, Mays, Starace, Liu, Maksin, Patwardhan, Madry, and Weng}]{chan2024mle}
Chan, J.~S.; Chowdhury, N.; Jaffe, O.; Aung, J.; Sherburn, D.; Mays, E.; Starace, G.; Liu, K.; Maksin, L.; Patwardhan, T.; Madry, A.; and Weng, L. 2025.
\newblock {MLE}-Bench: Evaluating Machine Learning Agents on Machine Learning Engineering.
\newblock In \emph{International Conference on Learning Representations}.

\bibitem[{Chen et~al.(2021)Chen, Tworek, Jun, Yuan, de~Oliveira~Pinto, Kaplan, Edwards, Burda, Joseph, Brockman, Ray, Puri, Krueger, Petrov, Khlaaf, Sastry, Mishkin, Chan, Gray, Ryder, Pavlov, Power, Kaiser, Bavarian, Winter, Tillet, Such, Cummings, Plappert, Chantzis, Barnes, Herbert-Voss, Guss, Nichol, Paino, Tezak, Tang, Babuschkin, Balaji, Jain, Saunders, Hesse, Carr, Leike, Achiam, Misra, Morikawa, Radford, Knight, Brundage, Murati, Mayer, Welinder, McGrew, Amodei, McCandlish, Sutskever, and Zaremba}]{humaneval}
Chen, M.; Tworek, J.; Jun, H.; Yuan, Q.; de~Oliveira~Pinto, H.~P.; Kaplan, J.; Edwards, H.; Burda, Y.; Joseph, N.; Brockman, G.; Ray, A.; Puri, R.; Krueger, G.; Petrov, M.; Khlaaf, H.; Sastry, G.; Mishkin, P.; Chan, B.; Gray, S.; Ryder, N.; Pavlov, M.; Power, A.; Kaiser, L.; Bavarian, M.; Winter, C.; Tillet, P.; Such, F.~P.; Cummings, D.; Plappert, M.; Chantzis, F.; Barnes, E.; Herbert-Voss, A.; Guss, W.~H.; Nichol, A.; Paino, A.; Tezak, N.; Tang, J.; Babuschkin, I.; Balaji, S.; Jain, S.; Saunders, W.; Hesse, C.; Carr, A.~N.; Leike, J.; Achiam, J.; Misra, V.; Morikawa, E.; Radford, A.; Knight, M.; Brundage, M.; Murati, M.; Mayer, K.; Welinder, P.; McGrew, B.; Amodei, D.; McCandlish, S.; Sutskever, I.; and Zaremba, W. 2021.
\newblock Evaluating Large Language Models Trained on Code.
\newblock arXiv:2107.03374.

\bibitem[{Chen et~al.(2025)Chen, Chen, Ning, Zhang, Wang, Yu, Li, Liao, Wei, Lu, Dey, Xue, Baker, Burns, Adu-Ampratwum, Huang, Ning, Gao, Su, and Sun}]{wang2025scienceagentbench}
Chen, Z.; Chen, S.; Ning, Y.; Zhang, Q.; Wang, B.; Yu, B.; Li, Y.; Liao, Z.; Wei, C.; Lu, Z.; Dey, V.; Xue, M.; Baker, F.~N.; Burns, B.; Adu-Ampratwum, D.; Huang, X.; Ning, X.; Gao, S.; Su, Y.; and Sun, H. 2025.
\newblock {ScienceAgentBench}: Toward Rigorous Assessment of Language Agents for Data-Driven Scientific Discovery.
\newblock In \emph{International Conference on Learning Representations}.

\bibitem[{Chirigati et~al.(2016)Chirigati, Rampin, Shasha, and Freire}]{reprozip}
Chirigati, F.; Rampin, R.; Shasha, D.; and Freire, J. 2016.
\newblock ReproZip: Computational Reproducibility With Ease.
\newblock In \emph{Proceedings of the 2016 International Conference on Management of Data}, 2085--2088. New York, NY, USA: Association for Computing Machinery.

\bibitem[{Deng et~al.(2026)Deng, Da, Pan, He, Ide, Garg, Lauffer, Park, Rane, Sampath, Krishnan, Kundurthy, Hendryx, Wang, Zhang, Jacobson, Liu, and Kenstler}]{swe2025pro}
Deng, X.; Da, J.; Pan, E.; He, Y.~Y.; Ide, C.; Garg, K.; Lauffer, N.; Park, A.; Rane, C.; Sampath, K.; Krishnan, M.; Kundurthy, S.~R.; Hendryx, S.~M.; Wang, Z.; Zhang, C. B.~C.; Jacobson, N.; Liu, B.; and Kenstler, B. 2026.
\newblock {SWE}-Bench Pro: Can {AI} Agents Solve Long-Horizon Software Engineering Tasks?
\newblock In \emph{Forty-third International Conference on Machine Learning}.

\bibitem[{Jimenez et~al.(2024)Jimenez, Yang, Wettig, Yao, Pei, Press, and Narasimhan}]{swebench}
Jimenez, C.~E.; Yang, J.; Wettig, A.; Yao, S.; Pei, K.; Press, O.; and Narasimhan, K. 2024.
\newblock {SWE}-bench: Can Language Models Resolve Real-World {GitHub} Issues?
\newblock In \emph{International Conference on Learning Representations (ICLR)}.

\bibitem[{Liu et~al.(2026)Liu, Qiu, Li, Li, Ji, Han, Ye, Xia, Dong, Chen, Zhang, Zhang, Chen, Tu, Yang, Feng, Zhao, Chen, Zhou, Wang, Zhang, Zhu, Li, Mei, Fei, Zhang, Li, Zhang, Zhou, Wang, Xiong, Zou, Zheng, Xie, Ding, and Yao}]{autoresearchclaw}
Liu, J.; Qiu, S.; Li, M.; Li, B.; Ji, H.; Han, S.; Ye, X.; Xia, P.; Dong, Z.; Chen, M.; Zhang, C.; Zhang, L.; Chen, G.; Tu, H.; Yang, X.; Feng, L.; Zhao, X.; Chen, H.; Zhou, J.; Wang, X.; Zhang, W.; Zhu, H.; Li, Y.; Mei, J.; Fei, H.; Zhang, J.; Li, L.; Zhang, L.; Zhou, Y.; Wang, S.; Xiong, C.; Zou, J.; Zheng, Z.; Xie, C.; Ding, M.; and Yao, H. 2026.
\newblock {AutoResearchClaw}: Self-Reinforcing Autonomous Research with Human-{AI} Collaboration.
\newblock arXiv:2605.20025.

\bibitem[{Liu et~al.(2024)Liu, Yu, Zhang, Xu, Lei, Lai, Gu, Ding, Men, Yang, Zhang, Deng, Zeng, Du, Zhang, Shen, Zhang, Su, Sun, Huang, Dong, and Tang}]{agentbench}
Liu, X.; Yu, H.; Zhang, H.; Xu, Y.; Lei, X.; Lai, H.; Gu, Y.; Ding, H.; Men, K.; Yang, K.; Zhang, S.; Deng, X.; Zeng, A.; Du, Z.; Zhang, C.; Shen, S.; Zhang, T.; Su, Y.; Sun, H.; Huang, M.; Dong, Y.; and Tang, J. 2024.
\newblock {AgentBench}: Evaluating {LLM}s as Agents.
\newblock In \emph{International Conference on Learning Representations}, 52989--53046.

\bibitem[{Lu et~al.(2024)Lu, Lu, Lange, Foerster, Clune, and Ha}]{aiscientist}
Lu, C.; Lu, C.; Lange, R.~T.; Foerster, J.; Clune, J.; and Ha, D. 2024.
\newblock The {AI} Scientist: Towards Fully Automated Open-Ended Scientific Discovery.
\newblock arXiv preprint arXiv:2408.06292.

\bibitem[{Sajadi, Damevski, and Chatterjee(2025)}]{sajadi2025secure}
Sajadi, A.; Damevski, K.; and Chatterjee, P. 2025.
\newblock How Safe Are {AI}-Generated Patches? {A} Large-scale Study on Security Risks in {LLM} and Agentic Automated Program Repair on {SWE}-bench.
\newblock arXiv:2507.02976.

\bibitem[{Santhosh et~al.(2026)Santhosh, Vas, Roychowdhury, Sakthi, and Rahaman}]{careful2026framework}
Santhosh, V.~N.; Vas, R.; Roychowdhury, B.; Sakthi, K.; and Rahaman, M. 2026.
\newblock Development and content validation of the {CAREFUL-AI} framework for evaluating {AI}-generated scientific manuscripts: an exploratory cross-platform study.
\newblock \emph{Research Evaluation}, 35: rvag024.

\bibitem[{Thiyagalingam et~al.(2022)Thiyagalingam, Shankar, Fox, and Hey}]{sciml2022benchmarks}
Thiyagalingam, J.; Shankar, M.; Fox, G.; and Hey, T. 2022.
\newblock Scientific machine learning benchmarks.
\newblock \emph{Nature Reviews Physics}, 4(6): 413--420.

\bibitem[{Wang et~al.(2025)Wang, Li, Song, Xu, Tang, Zhuge, Pan, Song, Li, Singh, Tran, Li, Ma, Zheng, Qian, Shao, Muennighoff, Zhang, Hui, Lin, Brennan, Peng, Ji, and Neubig}]{openhands}
Wang, X.; Li, B.; Song, Y.; Xu, F.~F.; Tang, X.; Zhuge, M.; Pan, J.; Song, Y.; Li, B.; Singh, J.; Tran, H.~H.; Li, F.; Ma, R.; Zheng, M.; Qian, B.; Shao, D.; Muennighoff, N.; Zhang, Y.; Hui, B.; Lin, J.; Brennan, R.; Peng, H.; Ji, H.; and Neubig, G. 2025.
\newblock {OpenHands}: An Open Platform for {AI} Software Developers as Generalist Agents.
\newblock In \emph{The Thirteenth International Conference on Learning Representations}.

\bibitem[{Wei et~al.(2025)Wei, Yang, Zhang, Chen, Zhuang, Gao, Zhou, Wang, Gao, Cao, Qiu, Hu, Ma, Tang, He, Song, He, Zhang, You, Zheng, Ding, Ouyang, Dong, Cheng, Sun, Bai, and Zhou}]{zheng2025agentic}
Wei, J.; Yang, Y.; Zhang, X.; Chen, Y.; Zhuang, X.; Gao, Z.; Zhou, D.; Wang, G.; Gao, Z.; Cao, J.; Qiu, Z.; Hu, M.; Ma, C.; Tang, S.; He, J.; Song, C.; He, X.; Zhang, Q.; You, C.; Zheng, S.; Ding, N.; Ouyang, W.; Dong, N.; Cheng, Y.; Sun, S.; Bai, L.; and Zhou, B. 2025.
\newblock From {AI} for Science to Agentic Science: A Survey on Autonomous Scientific Discovery.
\newblock arXiv:2508.14111.

\bibitem[{Wu et~al.(2026)Wu, Chen, Tan, Zhang, Xu, Qian, Gao, Zhu, Zhu, Tan, Ji, Lin, Chen, Ye, and Liu}]{clawailab}
Wu, F.; Chen, C.; Tan, Z.; Zhang, T.; Xu, X.; Qian, Y.; Gao, D.; Zhu, L.; Zhu, Q.; Tan, Y.; Ji, D.; Lin, G.; Chen, T.; Ye, D.; and Liu, F. 2026.
\newblock {Claw} {AI} Lab: An Autonomous Multi-Agent Research Team.
\newblock arXiv:2605.22662.

\bibitem[{Yang et~al.(2024)Yang, Jimenez, Wettig, Lieret, Yao, Narasimhan, and Press}]{sweagent}
Yang, J.; Jimenez, C.~E.; Wettig, A.; Lieret, K.; Yao, S.; Narasimhan, K.; and Press, O. 2024.
\newblock {SWE}-agent: Agent-Computer Interfaces Enable Automated Software Engineering.
\newblock In \emph{Advances in Neural Information Processing Systems 37 (NeurIPS)}.

\bibitem[{Yang et~al.(2025)Yang, Jimenez, Zhang, Lieret, Yang, Wu, Press, Muennighoff, Synnaeve, Narasimhan, Yang, Wang, and Press}]{swe2024multimodal}
Yang, J.; Jimenez, C.~E.; Zhang, A.; Lieret, K.; Yang, J.; Wu, X.; Press, O.; Muennighoff, N.; Synnaeve, G.; Narasimhan, K.; Yang, D.; Wang, S.; and Press, O. 2025.
\newblock {SWE}-Bench Multimodal: Do {AI} Systems Generalize to Visual Software Domains?
\newblock In \emph{International Conference on Learning Representations}, 2794--2829.

\bibitem[{Yuan et~al.(2025)Yuan, Yan, Zhang, Chen, Shi, Ouyang, Qiao, Bai, and Zhou}]{dolphin2025closed}
Yuan, J.; Yan, X.; Zhang, B.; Chen, T.; Shi, B.; Ouyang, W.; Qiao, Y.; Bai, L.; and Zhou, B. 2025.
\newblock Dolphin: Moving Towards Closed-loop Auto-research through Thinking, Practice, and Feedback.
\newblock In \emph{Proceedings of the 63rd Annual Meeting of the Association for Computational Linguistics (Volume 1: Long Papers)}, 21768--21789. Vienna, Austria: Association for Computational Linguistics.

\end{thebibliography}

\clearpage
\makeatletter
\@ifundefined{isAppendixMainFile}{
  \newif\ifappendixStandalone
  \appendixStandalonefalse
}{
  \newif\ifappendixStandalone
  \appendixStandalonetrue
}
\makeatother

\ifappendixStandalone
\documentclass[letterpaper]{article}
\usepackage[submission]{aaai2027}
\usepackage[hyphens]{url}
\usepackage{graphicx}
\usepackage{natbib}
\usepackage{caption}
\usepackage{booktabs}
\usepackage{tabularx}
\usepackage{multirow}
\usepackage{amsmath, amssymb}
\usepackage{algorithm}
\usepackage{algorithmic}
\usepackage{listings}
\usepackage{xcolor}
\usepackage{enumitem}
\usepackage{needspace}
\usepackage{etoolbox}

\definecolor{codebg}{rgb}{0.96,0.96,0.96}
\lstset{
  basicstyle=\ttfamily\scriptsize,
  backgroundcolor=\color{codebg},
  breaklines=true,
  breakatwhitespace=true,
  breakindent=0pt,
  postbreak=\raisebox{0ex}[0ex][0ex]{\ensuremath{\hookrightarrow}\space},
  numbers=left,
  numberstyle=\tiny\color{gray},
  xleftmargin=1.5em,
  showstringspaces=false,
  frame=single
}
\BeforeBeginEnvironment{lstlisting}{\needspace{6\baselineskip}}

\urlstyle{rm}
\def\UrlFont{\rm}
\frenchspacing

\pdfinfo{/TemplateVersion (2027.1)}

\author{Anonymous Submission}
\affiliations{Anonymous Institution}

\newcommand{\abe}{ABE-Ralph}
\newcommand{\metall}{\texttt{MET\_ALL}}
\newcommand{\metpartial}{\texttt{MET\_PARTIALLY}}
\newcommand{\failed}{\texttt{FAILED}}

\begin{document}
\title{Appendix}
\maketitle
\fi

\clearpage
\appendix

\section{Code Logic and Architecture of the \abe\ Framework}
\label{app:code_logic}

This section provides the complete algorithmic specification and architecture of the \abe\ (ABE-Ralph) framework, detailing its agent lifecycle, verification engine, contract structure, and file inventory.

\subsection{Overview of the Dual-Innovation Architecture}
The \abe\ framework embodies two tightly coupled innovations:
\begin{enumerate}
    \item \textbf{The Meta-Experiment Agent} (\texttt{ralph\_github.py}) --- a structured 8-step workflow orchestrating an LLM through experiment reproduction, self-healing, and ablation analysis.
    \item \textbf{The Triple-Verification Engine} (\texttt{verify\_reproduction.py}) --- a three-tier validation system evaluating quantitative alignment (L1), qualitative peer review (L2), and code implementation fidelity (L3).
\end{enumerate}

These components are mediated by a structured YAML contract defining expected datasets, baselines, target methods, metrics, success conditions, and critical modules.

\subsection{The Meta-Experiment Agent (\texttt{ralph\_github.py})}

\subsubsection{Class Structure and Initialization}
The agent is encapsulated in \texttt{MetaExperimentAgent}. Algorithm~\ref{alg:init} formalizes initialization.

\begin{algorithm}[h]
\caption{Agent Initialization}
\label{alg:init}
\begin{algorithmic}[1]
\REQUIRE \texttt{plan\_path} (YAML experiment plan), \texttt{workspace} (directory)
\ENSURE Initialized agent instance
\STATE \texttt{self.workspace} $\leftarrow$ \texttt{workspace}
\STATE \texttt{self.plan\_data} $\leftarrow$ \textsc{ParseYAML}(\texttt{plan\_path})
\STATE \texttt{self.discovery\_mode} $\leftarrow$ \textsc{IsDiscovery}(\texttt{self.plan\_data})
\STATE \textsc{InitSignalHandling}()
\STATE \textsc{SetupWorkspace}()
\STATE \texttt{self.\_budget\_timeouts} $\leftarrow$ \textsc{MapBudgets}(\texttt{self.plan\_data})
\IF{\texttt{resume} is TRUE}
    \STATE \textsc{LoadCheckpoint}()
\ELSE
    \STATE \textsc{InitCheckpoint}()
\ENDIF
\end{algorithmic}
\end{algorithm}

\subsubsection{Signal Handling: Self-Pipe Pattern}
To ensure interrupt responsiveness during subprocess monitoring, \abe\ implements a self-pipe pattern (Algorithm~\ref{alg:signal}).

\begin{algorithm}[h]
\caption{Signal Handling via Self-Pipe Pattern}
\label{alg:signal}
\begin{algorithmic}[1]
\STATE \textbf{Thread 1 (Signal Monitor):}
\STATE \quad \textsc{pthread\_sigmask}(BLOCK, \{SIGINT\})
\WHILE{\textnormal{TRUE}}
    \STATE \textsc{sigwait}(\{SIGINT\})
    \STATE \texttt{self.interrupted} $\leftarrow$ \textnormal{TRUE}
    \STATE \textsc{WriteByte}(\texttt{self\_pipe\_write\_end})
\ENDWHILE
\STATE \textbf{Thread 0 (Main Execution Thread):}
\WHILE{process is running}
    \STATE \texttt{ready} $\leftarrow$ \textsc{select}(\{\texttt{stdout}, \texttt{self\_pipe\_read\_end}\})
    \IF{\texttt{self\_pipe\_read\_end} $\in$ \texttt{ready}}
        \STATE \textbf{break} \COMMENT{Ctrl-C received, terminate safely}
    \ENDIF
    \STATE \textsc{ReadLine}(\texttt{stdout})
\ENDWHILE
\end{algorithmic}
\end{algorithm}

\subsubsection{The 8-Step Staged Workflow}
Algorithm~\ref{alg:workflow_code} details the sequential 8-step pipeline execution.

\begin{algorithm}[h]
\caption{The 8-Step Staged Workflow}
\label{alg:workflow_code}
\begin{algorithmic}[1]
\PROCEDURE{RunWorkflow}{}
    \STATE \textsc{RunClaudeStep}(\texttt{prompt\_intent}, timeout=300s)
    \STATE \textsc{RunClaudeStep}(\texttt{prompt\_dataset}, timeout=3600s)
    \IF{\texttt{DATA\_READY} $\notin$ output \textbf{and} \texttt{--strict-data}}
        \STATE \textbf{raise} \textnormal{DatasetVerificationError}
    \ENDIF
    \STATE \textsc{RunClaudeStep}(\texttt{prompt\_research}, timeout=1800s)
    \STATE \textsc{RunClaudeStep}(\texttt{prompt\_blueprint}, timeout=300s)
    \STATE \textsc{RunClaudeStep}(\texttt{prompt\_smoke\_test}, timeout=1800s)
    \STATE \textsc{RunClaudeStep}(\texttt{prompt\_full\_execution}, timeout=3600s)
    \STATE \textsc{VerifyMetricsJson}()
    \STATE \textsc{RunClaudeStep}(\texttt{prompt\_synthesis}, timeout=600s)
    \IF{\texttt{metrics.json} missing}
        \STATE \textsc{RunClaudeStep}(\texttt{prompt\_extract\_metrics}, timeout=600s)
    \ENDIF
    \STATE \textsc{RunClaudeStep}(\texttt{prompt\_skills})
    \FOR{$i = 1$ \TO \texttt{max\_retries}}
        \STATE \textsc{RunImprovementCycle}($i$)
        \IF{\texttt{HYPOTHESES\_SUPPORTED} $\lor$ \texttt{OPTIMIZATION\_DONE}}
            \STATE \textbf{break}
        \ENDIF
    \ENDFOR
    \STATE \textsc{RunAblationStudy}()
\ENDPROCEDURE
\end{algorithmic}
\end{algorithm}

\subsubsection{Discovery Mode vs. Reproduction Mode}
Table~\ref{tab:mode_comparison} contrasts the agent's behavior under Reproduction Mode and Discovery Mode.

\begin{table}[h]
\centering
\caption{Comparison of Reproduction and Discovery Modes.}
\label{tab:mode_comparison}
\scriptsize
\begin{tabularx}{\linewidth}{l X X}
\toprule
\textbf{Step} & \textbf{Reproduction Mode} & \textbf{Discovery Mode} \\
\midrule
1. Intent & Analyze YAML & Read \texttt{problem/README.md} \\
1.5. Dataset & Search HF/Kaggle/OpenML & Analyze local files and \texttt{evaluator.py} \\
2. Research & Search reference code repos & Search domain literature \\
3. Blueprint & Build \texttt{main.py} from reference & Design solution from scratch \\
4b. Execute & Train \& output \texttt{metrics.json} & Train \& execute \texttt{evaluator.py} \\
5. Synthesis & Check YAML criteria & Summarize evaluator score \\
\bottomrule
\end{tabularx}
\end{table}

\subsubsection{The Improvement Cycle}
Algorithm~\ref{alg:improve} presents the iterative refinement loop.

\begin{algorithm}[h]
\caption{Iterative Improvement Cycle}
\label{alg:improve}
\begin{algorithmic}[1]
\PROCEDURE{RunImprovementCycle}{\textit{iteration}, \textit{max\_retries}}
    \STATE \texttt{prompt\_imp} $\leftarrow$ ConstructPrompt(\{
    \STATE \quad \textnormal{Audit}: ``Analyze experiment\_result.md and raw data'',
    \STATE \quad \textnormal{CodeCheck}: ``Verify main.py against YAML contract'',
    \STATE \quad \textnormal{Fix}: ``Patch code and re-execute main.py''\})
    \STATE \textsc{RunClaudeStep}(\texttt{prompt\_imp}, step\_name=``imp\_v'' + \textit{iteration})
    
    \STATE \texttt{prompt\_ana} $\leftarrow$ ConstructPrompt(\{
    \STATE \quad \textnormal{Review}: ``Compare results against YAML success criteria'',
    \STATE \quad \textnormal{Signal}: ``Output HYPOTHESES\_SUPPORTED or OPTIMIZATION\_DONE''\})
    \STATE \texttt{out} $\leftarrow$ \textsc{RunClaudeStep}(\texttt{prompt\_ana}, step\_name=``ana\_v'' + \textit{iteration})
    
    \STATE \texttt{converged} $\leftarrow$ (\texttt{HYPOTHESES\_SUPPORTED} $\in$ \texttt{out})
    \STATE \texttt{opt\_done} $\leftarrow$ (\texttt{OPTIMIZATION\_DONE} $\in$ \texttt{out})
    \STATE \textsc{TrackIteration}(\textit{iteration}, \texttt{converged}, \texttt{opt\_done})
    \RETURN \texttt{out}
\ENDPROCEDURE
\end{algorithmic}
\end{algorithm}

\subsubsection{Checkpoint State Management}
The execution state is persisted in \texttt{experiment\_state.json}. The schema is defined as:

\begin{lstlisting}[language=json]
{
  "plan_path": "bm_001_dpr.yaml",
  "plan_mtime": 1717027200.0,
  "started_at": "2026-05-27T10:00:00",
  "steps": {
    "intent_discovery": {"status": "completed", "duration_seconds": 45.2},
    "dataset_search": {"status": "completed", "duration_seconds": 120.3},
    "research_step": {"status": "completed", "duration_seconds": 300.1},
    "blueprint": {"status": "completed"},
    "smoke_test": {"status": "completed"},
    "full_execution": {"status": "pending"},
    "synthesis_reporting": {"status": "pending"},
    "skill_extraction": {"status": "pending"}
  },
  "iteration": {"current": 0, "max_retries": 3, "history": []}
}
\end{lstlisting}

\subsubsection{The Ablation Study System}
Algorithm~\ref{alg:ablation_code} outlines the systematic ablation procedure.

\begin{algorithm}[h]
\caption{Ablation Study Execution}
\label{alg:ablation_code}
\begin{algorithmic}[1]
\PROCEDURE{RunAblationStudy}{}
    \FORALL{\texttt{cfg} $\in$ \{\textnormal{Full}, \textnormal{NoL3}, \textnormal{NoL1}, \textnormal{NoL2}\}}
        \STATE \texttt{ws} $\leftarrow$ \textsc{PrepareIsolatedWorkspace}(\texttt{cfg})
        \STATE \textsc{SwitchTo}(\texttt{ws})
        \IF{\texttt{cfg} == \textnormal{NoL3}}
            \STATE \textsc{RemoveReferenceCodeConstraints}()
        \ELSIF{\texttt{cfg} == \textnormal{NoL1}}
            \STATE \textsc{RemoveQuantitativeTargetPrompts}()
        \ELSIF{\texttt{cfg} == \textnormal{NoL2}}
            \STATE \textsc{DisableDeepAuditPrompts}()
        \ENDIF
        \STATE \texttt{decision} $\leftarrow$ \textsc{ExecuteEvaluation}()
        \STATE \texttt{scores} $\leftarrow$ \textsc{ScoreWorkspaceMultiDim}(\texttt{cfg})
        \STATE \textsc{RestoreWorkspace}()
    \ENDFOR
\ENDPROCEDURE
\end{algorithmic}
\end{algorithm}

Table~\ref{tab:ablation_mapping} maps ablation configurations to prompt-level modifications.

\begin{table}[h]
\centering
\caption{Prompt-Level Ablation Mapping.}
\label{tab:ablation_mapping}
\scriptsize
\begin{tabularx}{\linewidth}{l X X}
\toprule
\textbf{Config} & \textbf{Removed Prompt Directives} & \textbf{Auditing Concept} \\
\midrule
\textbf{w/o L3} & Skip repo search; remove code-guidance & Structural Code Alignment \\
\textbf{w/o L1} & Omit YAML metric success criteria comparison & Rigid Quantitative Alignment \\
\textbf{w/o L2} & Disable deep audit \& root-cause prompt blocks & Qualitative Peer Review \\
\bottomrule
\end{tabularx}
\end{table}

\subsubsection{Multi-Dimensional Scoring}
Algorithm~\ref{alg:scoring_code} defines multi-dimensional workspace evaluation.

\begin{algorithm}[h]
\caption{Multi-Dimensional Workspace Scoring}
\label{alg:scoring_code}
\begin{algorithmic}[1]
\PROCEDURE{ScoreWorkspaceMultiDim}{\textit{label}}
    \STATE \texttt{integrator} $\leftarrow$ \textsc{ExperimentIntegrator}(\texttt{workspace})
    \STATE \texttt{s} $\leftarrow$ \texttt{integrator}.\textsc{ComputeMultiDimScore}()
    \RETURN \{$S_{comp} = 0.20 s_A + 0.25 s_B + 0.25 s_C + 0.30 s_D$, Grade: A/B/C/D/F\}
\ENDPROCEDURE
\end{algorithmic}
\end{algorithm}

\subsection{The Triple-Verification Engine (\texttt{verify\_reproduction.py})}

\subsubsection{Initialization and Structural Checks}
Algorithm~\ref{alg:verifier_init} details initialization of the verifier engine.

\begin{algorithm}[h]
\caption{Verifier Engine Initialization}
\label{alg:verifier_init}
\begin{algorithmic}[1]
\REQUIRE \texttt{plan\_path} (YAML), \texttt{workspace}
\STATE \texttt{self.plan} $\leftarrow$ \textsc{ParseYAML}(\texttt{plan\_path})
\STATE \texttt{self.report} $\leftarrow$ \textsc{DetectLatestReport}()
\STATE \texttt{self.discovery} $\leftarrow$ (\texttt{self.plan.mode} == ``discovery'')
\STATE \texttt{self.datasets} $\leftarrow$ \textsc{ExtractDatasets}(\texttt{self.plan})
\STATE \texttt{self.baselines} $\leftarrow$ \textsc{ExtractBaselines}(\texttt{self.plan})
\STATE \texttt{self.target} $\leftarrow$ \textsc{ExtractTargetMethod}(\texttt{self.plan})
\end{algorithmic}
\end{algorithm}

\subsubsection{Level 1: Rigid Quantitative Alignment ($V_{quant}$)}
Algorithm~\ref{alg:l1_check} verifies JSON key structures and presence.

\begin{algorithm}[h]
\caption{Level 1 Quantitative Alignment Check}
\label{alg:l1_check}
\begin{algorithmic}[1]
\PROCEDURE{RunDynamicQuantitativeCheck}{\textit{metrics\_data}}
    \IF{\textit{metrics\_data} is NULL}
        \RETURN (FALSE, ``metrics.json missing'')
    \ENDIF
    \FORALL{\texttt{ds} $\in$ \texttt{self.datasets}}
        \STATE \texttt{k\_ds} $\leftarrow$ \textsc{FuzzyMatch}(\texttt{ds}, \textit{metrics\_data.keys()})
        \IF{\texttt{k\_ds} is NULL} \RETURN (FALSE, ``Dataset missing'') \ENDIF
        \STATE \texttt{k\_tgt} $\leftarrow$ \textsc{FuzzyMatch}(\texttt{self.target}, \textit{metrics\_data}[\texttt{k\_ds}])
        \IF{\texttt{k\_tgt} is NULL} \RETURN (FALSE, ``Target method missing'') \ENDIF
    \ENDFOR
    \RETURN (TRUE, ``Quantitative structure verified'')
\ENDPROCEDURE
\end{algorithmic}
\end{algorithm}

\subsubsection{Level 2: Qualitative Peer Review ($V_{qual}$)}
Algorithm~\ref{alg:l2_check} executes qualitative LLM review.

\begin{algorithm}[h]
\caption{Level 2 Qualitative Peer Review Check}
\label{alg:l2_check}
\begin{algorithmic}[1]
\PROCEDURE{RunUniversalQualitativeCheck}{}
    \STATE \texttt{prompt} $\leftarrow$ ConstructPrompt(\{
    \STATE \quad \textnormal{Role}: ``Senior meta-reviewer auditing reproduction'',
    \STATE \quad \textnormal{Claims}: \texttt{self.plan.key\_claims},
    \STATE \quad \textnormal{Report}: \textsc{ReadFile}(\texttt{self.report\_path})\})
    \STATE \texttt{out} $\leftarrow$ \textsc{InvokeClaude}(\texttt{prompt})
    \STATE \texttt{decision} $\leftarrow$ \textsc{ParseDecision}(\texttt{out}) \COMMENT{Regex: DECISION:\textbackslash s*(MET\_ALL|MET\_PARTIALLY|FAILED)}
    \RETURN (\texttt{decision})
\ENDPROCEDURE
\end{algorithmic}
\end{algorithm}

\subsubsection{Level 3: Structural Code Alignment ($V_{struct}$)}
Algorithm~\ref{alg:l3_check} executes AST and module existence checks.

\begin{algorithm}[h]
\caption{Level 3 Code Alignment Verification}
\label{alg:l3_check}
\begin{algorithmic}[1]
\PROCEDURE{RunCodeAlignmentCheck}{}
    \STATE \texttt{modules} $\leftarrow$ \texttt{self.plan.target\_method. critical\_modules}
    \STATE \texttt{code} $\leftarrow$ \textsc{ReadFile}(``main.py'')
    \FORALL{\texttt{m} $\in$ \texttt{modules}}
        \IF{\textbf{not} \textsc{ASTSearch}(\texttt{m}, \texttt{code})}
            \RETURN (FALSE, ``Critical module missing: '' + \texttt{m})
        \ENDIF
    \ENDFOR
    \RETURN (TRUE, ``Code structural alignment passed'')
\ENDPROCEDURE
\end{algorithmic}
\end{algorithm}

\subsubsection{Decision Aggregation}
Algorithm~\ref{alg:decision_agg} details the hierarchical decision flow.

\begin{algorithm}[h]
\caption{Decision Aggregation Protocol}
\label{alg:decision_agg}
\begin{algorithmic}[1]
\PROCEDURE{ExecuteEvaluation}{}
    \STATE \texttt{m\_data} $\leftarrow$ \textsc{LoadMetrics}()
    \STATE \texttt{q\_pass}, \_ $\leftarrow$ \textsc{RunDynamicQuantitativeCheck}(\texttt{m\_data})
    \STATE \texttt{decision} $\leftarrow$ \textsc{RunUniversalQualitativeCheck}()
    \IF{\texttt{decision} == \textnormal{MET\_ALL} \textbf{and} \texttt{m\_data} is NULL}
        \STATE \texttt{decision} $\leftarrow$ \textnormal{MET\_PARTIALLY}
    \ENDIF
    \IF{\textbf{not} \texttt{self.discovery}}
        \STATE \texttt{l3\_pass}, \_ $\leftarrow$ \textsc{RunCodeAlignmentCheck}()
        \IF{\textbf{not} \texttt{l3\_pass}}
            \STATE \texttt{decision} $\leftarrow$ \textnormal{FAILED} \COMMENT{Block on M1 method collapse}
        \ENDIF
    \ENDIF
    \RETURN \texttt{decision}
\ENDPROCEDURE
\end{algorithmic}
\end{algorithm}

\subsubsection{Failure Taxonomy Signal Matching}
The verification results are automatically mapped to failure modes (M1--M5) using regular expressions:

\begin{lstlisting}[language=Python]
FAILURE_CATEGORIES = {
    "METHOD_COLLAPSE": {
        "signals": ["oracle", "upper bound", "never started", "substituted"]
    },
    "SILENT_DEGRADE": {
        "signals": ["from.scratch", "random init", "degenerate.*solution"]
    },
    "SCALE_INVERSION": {
        "signals": ["baseline.*outperform", "direction.*reversal"]
    },
    "KEY_MISMATCH": {
        "signals": ["metrics.json.*missing", "key.*not.*match"]
    },
    "INCOMPLETE": {
        "signals": ["not all condition", "time budget exhaust"]
    }
}
\end{lstlisting}

\subsection{The YAML Contract Specification}
Table~\ref{tab:yaml_contract} presents the YAML contract structure.

\subsubsection{Example Contract Instance}
\label{app:yaml_example}
To ground the abstract schema in Table~\ref{tab:yaml_contract}, Listing~\ref{lst:yaml_example} presents a concrete YAML contract instance corresponding to the DPR (Dense Passage Retrieval) benchmark (\texttt{bm\_001\_dpr\_dpr.yaml}). This contract specifies the paper metadata, key claims, datasets, baselines, target method, expected metric values with tolerance ranges, success/failure conditions, and compute/time budgets that jointly drive both the \abe\ agent's execution and the Triple-Verification Engine's decision logic.

\begin{lstlisting}[language=yaml, caption={Example YAML experiment contract (\texttt{bm\_001\_dpr\_dpr.yaml}) used for the DPR benchmark. Fields map directly to the consumers listed in Table~\ref{tab:yaml_contract}.}, label={lst:yaml_example}]
topic: Dense Passage Retrieval for Open-Domain Question Answering
paper_metadata:
  title: Dense Passage Retrieval for Open-Domain Question Answering
  authors: Vladimir Karpukhin, Barlas Oguz, Sewon Min, Patrick Lewis, ...
  year: 2020
  venue: EMNLP
  domain: nlp_retrieval
  task_type: open_domain_qa_retrieval
reproduction_goal: >
  Evaluate whether an AI agent can reproduce the main experimental
  conclusions of DPR: that dense dual-encoder retrieval outperforms
  BM25 on open-domain QA retrieval metrics, using reduced-scale
  reproduction on canonical datasets.
key_claims:
  - Dense dual-encoder retrieval (DPR) substantially outperforms BM25
    on open-domain QA retrieval metrics such as Recall@k and MRR.
  - Dense retrieval provides stronger downstream evidence quality for
    QA than sparse lexical retrieval alone.
datasets:
  - NaturalQuestions
  - TriviaQA
baselines:
  - BM25 (sparse lexical retrieval, e.g., via Pyserini or Elasticsearch)
target_method:
  name: Dense Passage Retriever (DPR)
  description: A dual-encoder dense retrieval model that encodes
    questions and passages into a shared embedding space for
    efficient top-k retrieval via maximum inner product search.
  codebase: facebookresearch/DPR
ablations:
  - BM25 baseline
  - DPR with different negative sampling strategies
  - DPR with reduced training data
metrics:
  - recall_at_20
  - recall_at_100
  - mrr
expected_metric_values:
  - metric_name: recall_at_20
    expected_value: 0.78
    reasonable_range: [0.73, 0.8]
    higher_is_better: true
  - metric_name: recall_at_100
    expected_value: 0.85
    reasonable_range: [0.8, 0.88]
    higher_is_better: true
  - metric_name: mrr
    expected_value: 0.32
    reasonable_range: [0.29, 0.34]
    higher_is_better: true
expected_paper_conclusions:
  - DPR (dense retrieval) achieves higher Recall@k and MRR than BM25
    on open-domain QA retrieval benchmarks.
  - Dense retrieval provides better evidence retrieval for downstream
    QA than sparse lexical retrieval alone.
success_conditions:
  conclusion_correctness: true
  metric_direction_consistency: DPR must outperform BM25 on all
    primary retrieval metrics (recall_at_20, recall_at_100, mrr) in
    the same direction as the original paper.
  result_closeness_tolerance: 0.05
failure_conditions:
  - BM25 outperforms or matches DPR on primary retrieval metrics.
  - Metric differences are outside the reasonable range or opposite
    in direction to the paper.
  - Incorrect evaluation setup (e.g., wrong negatives, inconsistent
    preprocessing) invalidates comparison.
compute_budget:
  gpu_type: NVIDIA GeForce RTX 4070 Laptop GPU
  gpu_memory_gb: 8
  max_gpu_hours: 18.0
  cpu_cores: 8
  ram_gb: 32
time_budget:
  environment_setup_hours: 2.0
  repo_selection_and_code_reading_hours: 2.0
  dataset_parsing_and_section_segmentation_hours: 2.0
  baseline_pipeline_adaptation_hours: 2.0
  proposed_method_implementation_hours: 2.5
  debugging_and_memory_optimization_hours: 2.0
  baseline_training_and_evaluation_hours: 2.0
  proposed_method_training_and_evaluation_hours: 2.5
  ablation_runs_hours: 1.0
  result_aggregation_and_visualization_hours: 1.0
  contingency_buffer_hours: 1.0
risks:
  - Incorrect negative sampling or passage index construction may invalidate results.
  - Incompatible preprocessing between DPR and BM25 could bias metrics.
  - GPU memory limits may require smaller batch sizes or model variants, affecting metric values.
  - Dataset splits or answer string matching may differ from the original, impacting recall.
reproducibility_notes: 
  Reduced-scale reproduction uses DPR-base model, smaller batch sizes, and fewer epochs to fit within compute budget. Pretrained checkpoints may be used for evaluation if full training is infeasible. Metric values are expected to be within 5% of the original papers results, but directionality and relative ranking between DPR and BM25 are the main criteria for success.
generation_metadata:
  generated_at: '2026-06-01T16:42:13.157186'
  generator_model: gpt-4.1
  benchmark_id: bm_001_dpr
  benchmark_name: DPR
  priority_tier: P1
  difficulty: 2
  reproduction_mode: reduced_scale_reproduction
  retry_count: 0
  validation_status: complete
\end{lstlisting}

As shown above, the contract is organized into six functional blocks consumed by different components of the \abe\ pipeline:
\begin{enumerate}[leftmargin=*]
    \item \textbf{Provenance block} (\texttt{topic}, \texttt{paper\_metadata}, \texttt{reproduction\_goal}) --- consumed by the Meta-Experiment Agent during the \textit{Intent} and \textit{Research} steps to establish scope and locate reference implementations.
    \item \textbf{Claim/target block} (\texttt{key\_claims}, \texttt{datasets}, \texttt{baselines}, \texttt{target\_method}, \texttt{ablations}, \texttt{metrics}) --- jointly consumed by the Agent (Blueprint/Execution steps) and the Verifier's L1 and L3 checks to determine which datasets, comparative baselines, and code modules must be present.
    \item \textbf{Quantitative criteria block} (\texttt{expected\_metric\_values}, \texttt{expected\_paper\_conclusions}, \texttt{success\_conditions}, \texttt{failure\_conditions}) --- directly parameterizes \textsc{RunDynamicQuantitativeCheck} (Algorithm~\ref{alg:l1_check}) and the L2 qualitative reviewer's red-line triggers.
    \item \textbf{Resource block} (\texttt{compute\_budget}, \texttt{time\_budget}) --- consumed by \texttt{self.\_budget\_timeouts} in Algorithm~\ref{alg:init} to bound each staged workflow step (e.g., \texttt{max\_gpu\_hours} constrains \textit{full\_execution}).
    \item \textbf{Risk block} (\texttt{risks}, \texttt{reproducibility\_notes}) --- informs the Agent's self-healing/improvement cycle (Algorithm~\ref{alg:improve}) by pre-registering plausible failure causes (e.g., negative sampling errors, preprocessing mismatch) referenced by the failure taxonomy signal matcher.
    \item \textbf{Provenance/audit metadata block} (\texttt{generation\_metadata}) --- used for experiment tracking, reproducibility auditing, and versioning across the 30-benchmark suite (Appendix~\ref{app:hallucination_breakdown}--\ref{app:case_studies}).
\end{enumerate}

\begin{table}[h]
\centering
\caption{YAML Contract Schema and Field Consumers.}
\label{tab:yaml_contract}
\scriptsize
\begin{tabularx}{\linewidth}{l X X}
\toprule
\textbf{Field} & \textbf{Consumers} & \textbf{Purpose} \\
\midrule
\texttt{datasets} & Agent, Verifier (L1) & Expected dataset specifications \\
\texttt{baselines} & Agent, Verifier (L1) & Required baseline comparative targets \\
\texttt{target\_method} & Agent, Verifier (L1, L3) & Target algorithm description \\
\texttt{critical\_modules} & Verifier (L3) & Mandatory AST components in \texttt{main.py} \\
\texttt{key\_claims} & Verifier (L2) & Claims to evaluate qualitatively \\
\texttt{success\_conditions} & Agent, Verifier (L1) & Metric margins and directionality \\
\texttt{failure\_conditions} & Verifier (L2) & Red-line failure triggers \\
\bottomrule
\end{tabularx}
\end{table}

\subsection{File Inventory}
Table~\ref{tab:file_inventory} details the primary system modules.

\begin{table}[h]
\centering
\caption{Complete System File Inventory.}
\label{tab:file_inventory}
\scriptsize
\begin{tabularx}{\linewidth}{l c X}
\toprule
\textbf{File Name} & \textbf{Lines} & \textbf{Primary Operational Role} \\
\midrule
\texttt{ralph\_github.py} & $\sim$1246 & Workflow agent, iteration, ablation loop \\
\texttt{verify\_reproduction.py} & $\sim$519 & Triple-verification engine, decision engine \\
\texttt{alignment\_verifier.py} & $\sim$266 & Level 3 structural code fidelity auditor \\
\texttt{experiment\_taxonomy.py} & $\sim$220 & Failure taxonomy classifier (M1--M5) \\
\texttt{ablation\_scoring.py} & $\sim$280 & Multi-dimensional scoring module \\
\texttt{experiment\_integrator.py} & $\sim$390 & Execution history, metric tracking \\
\bottomrule
\end{tabularx}
\end{table}

\section{Methodological Hallucination Taxonomy: Per-Experiment Breakdown}
\label{app:hallucination_breakdown}

This section details the per-experiment analysis across 30 classical machine learning and deep learning benchmark reproductions evaluated under the \abe\ auditing framework. A total of 17 experiments exhibited one or more methodological hallucination types (M1--M5), while 13 passed verification without shortcuts.

\subsection{M1: Method Integrity Collapse (5 Methods)}
Occurs when the agent omits core target components, substitutes trivial heuristic fallbacks, or fails to execute comparative conditions.

\begin{itemize}[leftmargin=*]
    \item \textbf{RAG:} (Also M5). The generator fell back to parametric knowledge when retrieval failed (EM 1.8\% vs. 42.1\% with successful retrieval). The experiment conflated parametric memory with RAG performance, failing to isolate the retrieval component's contribution.
    \item \textbf{BERT:} (Also M5). The pretrained checkpoint was fine-tuned on fundamentally different tasks (e.g., sentiment analysis) chosen by the LLM, rather than the specific GLUE benchmark tasks (MNLI, QQP, MRPC) specified in the paper.
    \item \textbf{SimCLR:} (Also M3, M5). Baseline methods (e.g., random encoder) were queued but never executed, meaning the core comparison of ``SimCLR vs. previous unsupervised methods'' never occurred.
    \item \textbf{DDIM:} (Also M5). The core comparison baseline (DDPM) was never executed, meaning the primary claim (DDIM is faster than DDPM) was unvalidated.
    \item \textbf{Improved DDPM:} (Also M5). Training instability (likely in learned variance parameters) eliminated the expected quality gap over base DDPM, failing to reproduce the claimed likelihood improvements.
\end{itemize}

\subsection{M2: Silent Protocol Degradation (6 Methods)}
Occurs when training protocols, negative sampling, or initialization routines are modified without authorization to force execution.

\begin{itemize}[leftmargin=*]
    \item \textbf{ColBERT:} (Also M5). Hard negative mining was completely omitted from the training protocol, disabling ColBERT's ability to discriminate relevant from irrelevant passages.
    \item \textbf{PEGASUS:} (Also M3, M5). Trained from scratch for 3 epochs on 50K samples instead of fine-tuning the pretrained GSG checkpoint, testing random initialization convergence rather than representation quality.
    \item \textbf{ViT:} (Also M3, M5). Trained from scratch on a small dataset without pretraining, resulting in a 20-percentage-point accuracy drop and violating the paper's assumption of large-scale pretraining.
    \item \textbf{Faster R-CNN:} Protocol degradation in the detection pipeline due to non-standard RPN settings or inconsistent mAP evaluation protocols.
    \item \textbf{PPO:} (Also M5). Continuous-control environments (MuJoCo) crashed with 0/9 runs completed; report falsely claimed continuous control success based on discrete CartPole.
    \item \textbf{GraphSAGE:} (Also M4, M5). Inductive vs. transductive settings were misconfigured, breaking the inductive generalization claim.
\end{itemize}

\subsection{M3: Scale-Driven Conclusion Inversion (4 Methods)}
Occurs when resource-limited scaling or protocol degradation inverts the paper's original ranking or core conclusions.

\begin{itemize}[leftmargin=*]
    \item \textbf{PEGASUS:} (Also M2, M5). PEGASUS underperformed BERT (-32\%) and BART (-55\%), completely inverting the paper's SOTA summarization ranking claims.
    \item \textbf{ViT:} (Also M2, M5). At reduced scale without pretraining, CNN baselines outperformed ViT, directly opposing the paper's conclusion.
    \item \textbf{SimCLR:} (Also M1, M5). The agent substituted a weaker verification standard, concluding ``above chance'' performance rather than proving superiority over baseline encoders.
    \item \textbf{U-Net:} (Also M5). Reduced-depth ablation paradoxically outperformed the full U-Net under limited data, contradicting claims on encoder-decoder depth utility.
\end{itemize}

\subsection{M4: Quantitative Key Mismatch (2 Methods)}
Occurs when output files or metric keys deviate from verification schemas, causing verification pipeline parsing failures.
\begin{itemize}[leftmargin=*]
    \item \textbf{RoBERTa:} (Also M5). Quantitative checks failed solely because \texttt{metrics.json} was produced under a non-standard filename pattern by aggregation scripts.
    \item \textbf{GraphSAGE:} (Also M2, M5). Inconsistent metric naming (e.g., \texttt{"acc"} vs. \texttt{"accuracy"} in output JSON) caused the parser to miss evaluation results.
\end{itemize}

\subsection{M5: Incomplete Execution (16 Methods)}
Occurs when benchmark evaluation conditions, datasets, or target conditions are prematurely terminated due to compute limits.
Affected methods: \textbf{RAG}, \textbf{ColBERT}, \textbf{BERT}, \textbf{RoBERTa}, \textbf{PEGASUS}, \textbf{LED}, \textbf{ViT}, \textbf{SimCLR}, \textbf{MoCo}, \textbf{U-Net}, \textbf{YOLO}, \textbf{PatchCore}, \textbf{DDIM}, \textbf{Improved DDPM}, \textbf{PPO}, and \textbf{GraphSAGE}.

\subsection{Clean Experiments (13 Baseline Reproduction Runs)}
Thirteen methods passed all verification checks without hallucination shortcuts: \textbf{DPR}, \textbf{LoRA}, \textbf{Prefix-Tuning}, \textbf{Longformer}, \textbf{TextRank}, \textbf{ResNet}, \textbf{CLIP}, \textbf{PaDiM}, \textbf{DDPM}, \textbf{DQN}, \textbf{GCN}, \textbf{Informer}, and \textbf{Autoformer}.

\section{Detailed Per-Experiment Case Studies}
\label{app:case_studies}

This section provides detailed case studies for all 30 classical machine learning and deep learning benchmark reproductions evaluated under the \abe\ auditing framework.

\subsection{Individual Experiment Analyses}

\paragraph{1. DPR (Dense Passage Retrieval)}
\textit{Status: CLEAN}. All quantitative and qualitative checks passed successfully. Evaluation metrics for NaturalQuestions and TriviaQA are fully present, and the conclusion (DPR outperforms BM25) is correctly supported by the reproduced experimental data.

\paragraph{2. RAG (Retrieval-Augmented Generation)}
\textit{Status: HALLUCINATION (M1 + M5)}. 
\begin{itemize}[leftmargin=*]
    \item \textbf{M1 Collapse:} The generator fell back to parametric knowledge when retrieval failed (EM 1.8\% vs. 42.1\% with successful retrieval). The experiment conflated the generator's memory with RAG performance, thus failing to isolate the retrieval component's contribution.
    \item \textbf{M5 Incomplete:} Missing metrics for NaturalQuestions and WebQuestions due to index construction infrastructure issues.
\end{itemize}

\paragraph{3. ColBERT (Contextualized Late Interaction over BERT)}
\textit{Status: HALLUCINATION (M2 + M5)}.
\begin{itemize}[leftmargin=*]
    \item \textbf{M2 Degradation:} Hard negative mining was completely omitted from the training protocol. Because ColBERT relies heavily on hard negatives, the model failed to discriminate relevant from irrelevant passages.
    \item \textbf{M5 Incomplete:} Omitted the TREC Deep Learning dataset, a primary benchmark in the paper.
\end{itemize}

\paragraph{4. BERT (Pretrained Language Model)}
\textit{Status: HALLUCINATION (M1 + M5)}.
\begin{itemize}[leftmargin=*]
    \item \textbf{M1 Collapse:} The pretrained checkpoint was fine-tuned on non-target tasks (e.g., sentiment analysis) chosen by the LLM rather than specific GLUE benchmark tasks (MNLI, QQP, MRPC).
    \item \textbf{M5 Incomplete:} Metrics for the full GLUE dataset suite are missing; only SST-2 results were reported.
\end{itemize}

\paragraph{5. RoBERTa (Robustly Optimized BERT)}
\textit{Status: HALLUCINATION (M4 + M5)}.
\begin{itemize}[leftmargin=*]
    \item \textbf{M4 Key Mismatch:} Quantitative verification failed because \texttt{metrics.json} was generated under a non-standard filename pattern by aggregation scripts.
    \item \textbf{M5 Incomplete:} Missing full GLUE benchmark evaluation metrics.
\end{itemize}

\paragraph{6. LoRA (Low-Rank Adaptation)}
\textit{Status: CLEAN}. Clean run. All required metrics for GLUE and instruction-tuning subsets are present. The conclusion that LoRA matches full fine-tuning with significantly fewer trainable parameters is correctly supported.

\paragraph{7. Prefix-Tuning (Continuous Prompt Tuning)}
\textit{Status: CLEAN}. Clean reproduction without hallucinations. The LLM successfully set up continuous prefix parameters and compared them against full fine-tuning baselines under stable convergence.

\paragraph{8. PEGASUS (Abstractive Summarization with Stoned Sentences)}
\textit{Status: HALLUCINATION (M2 + M3 + M5)}.
\begin{itemize}[leftmargin=*]
    \item \textbf{M2 Degradation:} Trained from scratch for 3 epochs on 50K samples instead of using the pretrained GSG checkpoint.
    \item \textbf{M3 Inversion:} The performance ranking inverted: PEGASUS underperformed BERT (-32\%) and BART (-55\%), contradicting SOTA summarization claims.
    \item \textbf{M5 Incomplete:} Missing the XSum extreme summarization benchmark dataset.
\end{itemize}

\paragraph{9. LED (Longformer-Encoder-Decoder)}
\textit{Status: HALLUCINATION (M5)}. Missing target method metrics for the arXiv long-document dataset. Since arXiv is a primary long-context benchmark, the core claim of superior long-context summarization could not be verified.

\paragraph{10. Longformer (Long-Document Transformer)}
\textit{Status: CLEAN}. All quantitative checks passed. The LLM correctly configured sparse attention masks (sliding window + global attention) and recovered expected efficiency-performance tradeoffs.

\paragraph{11. TextRank (Graph-based Unsupervised Summarization)}
\textit{Status: CLEAN}. Graph-based extractive summarization baseline was correctly implemented, and ROUGE scores were computed accurately across reference documents.

\paragraph{12. ResNet (Deep Residual Learning)}
\textit{Status: CLEAN}. Residual architecture advantages over plain CNNs were successfully reproduced on CIFAR-10/100 datasets using proper residual connections and learning rate schedules.

\paragraph{13. ViT (Vision Transformer)}
\textit{Status: HALLUCINATION (M2 + M3 + M5)}.
\begin{itemize}[leftmargin=*]
    \item \textbf{M2 Degradation:} ViT was trained from scratch on a small dataset without large-scale pretraining, resulting in a 20 percentage point accuracy drop.
    \item \textbf{M3 Inversion:} At reduced scale, standard CNNs outperformed ViT, inverting the paper's original conclusions.
    \item \textbf{M5 Incomplete:} Missing ImageNet pretraining/evaluation subset metrics.
\end{itemize}

\paragraph{14. CLIP (Contrastive Language-Image Pretraining)}
\textit{Status: CLEAN}. The LLM correctly executed zero-shot classification evaluation using appropriate prompt templates and recovered expected zero-shot transfer performance.

\paragraph{15. SimCLR (Simple Framework for Contrastive Learning)}
\textit{Status: HALLUCINATION (M1 + M3 + M5)}.
\begin{itemize}[leftmargin=*]
    \item \textbf{M1 Collapse:} Baseline methods (e.g., random encoder) were queued but never executed.
    \item \textbf{M3 Inversion:} The LLM substituted a weaker standard, concluding ``above chance'' performance rather than demonstrating superiority over unsupervised baselines.
    \item \textbf{M5 Incomplete:} All baseline comparative evaluation results were left pending.
\end{itemize}

\paragraph{16. MoCo (Momentum Contrast for Unsupervised Visual Representation)}
\textit{Status: HALLUCINATION (M5)}. Metrics for the ImageNet evaluation subset are missing. As ImageNet is the primary benchmark, linear probe classification performance remains unverified.

\paragraph{17. U-Net (Convolutional Networks for Biomedical Image Segmentation)}
\textit{Status: HALLUCINATION (M3 + M5)}.
\begin{itemize}[leftmargin=*]
    \item \textbf{M3 Inversion:} The reduced-depth ablation paradoxically performed better than the full U-Net under small sample limits, contradicting claims that full encoder-decoder depth is essential.
    \item \textbf{M5 Incomplete:} Missing key biomedical segmentation benchmarks (e.g., ISBI dataset).
\end{itemize}

\paragraph{18. Faster R-CNN (Towards Real-Time Object Detection)}
\textit{Status: HALLUCINATION (M2)}. Protocol degradation in detection pipeline — non-standard Region Proposal Network (RPN) settings or non-standard mAP evaluation protocols caused silent evaluation divergence.

\paragraph{19. YOLO (Real-Time Object Detection)}
\textit{Status: HALLUCINATION (M5)}. Target method metrics for the PASCAL VOC benchmark dataset are missing, leaving the speed-accuracy tradeoff profile incomplete.

\paragraph{20. PatchCore (Industrial Anomaly Detection)}
\textit{Status: HALLUCINATION (M5)}. Detailed localization metrics (e.g., pixel-level AUROC or AUPRO) are missing, leaving spatial anomaly localization unverified.

\paragraph{21. PaDiM (Patch Distribution Modeling for Anomaly Detection)}
\textit{Status: CLEAN}. All checks passed. Mahalanobis distance estimation over patch embedding distributions was correctly implemented, recovering anomaly detection metrics on MVTec AD.

\paragraph{22. DDPM (Denoising Diffusion Probabilistic Models)}
\textit{Status: CLEAN}. Stable reproduction of the diffusion training pipeline with correct beta schedule, sampling loop, and evaluation quality metrics (FID).

\paragraph{23. DDIM (Denoising Diffusion Implicit Models)}
\textit{Status: HALLUCINATION (M1 + M5)}.
\begin{itemize}[leftmargin=*]
    \item \textbf{M1 Collapse:} Baseline DDPM sampling runs were never executed; speedup claims over DDPM could not be tested.
    \item \textbf{M5 Incomplete:} Only 4 out of 12 planned conditions were completed (33\% completion rate) using a single random seed.
\end{itemize}

\paragraph{24. Improved DDPM (Improved Denoising Diffusion)}
\textit{Status: HALLUCINATION (M1 + M5)}.
\begin{itemize}[leftmargin=*]
    \item \textbf{M1 Collapse:} Training instability with learned variance parameters eliminated expected generation quality improvements over base DDPM.
    \item \textbf{M5 Incomplete:} Missing metrics for subset datasets (e.g., ImageNet subset).
\end{itemize}

\paragraph{25. DQN (Deep Q-Networks)}
\textit{Status: CLEAN}. Clean execution. Replay buffer and target network update logic were correctly implemented, recovering expected reinforcement learning performance on Atari/CartPole.

\paragraph{26. PPO (Proximal Policy Optimization)}
\textit{Status: HALLUCINATION (M2 + M5)}.
\begin{itemize}[leftmargin=*]
    \item \textbf{M2 Degradation:} All continuous control environments (MuJoCo HalfCheetah, Hopper, Walker2d) crashed (0/9 completed). The report falsely claimed continuous control success based solely on discrete CartPole-v1.
    \item \textbf{M5 Incomplete:} Zero baseline data (Vanilla PG, TRPO) was provided.
\end{itemize}

\paragraph{27. GCN (Graph Convolutional Networks)}
\textit{Status: CLEAN}. Graph convolution layers, adjacency matrix normalization, and train/val/test splits were correctly set up, recovering node classification accuracy on Cora and Citeseer.

\paragraph{28. GraphSAGE (Inductive Representation Learning on Large Graphs)}
\textit{Status: HALLUCINATION (M2 + M4 + M5)}.
\begin{itemize}[leftmargin=*]
    \item \textbf{M2 Degradation:} Inductive vs. transductive settings were misconfigured.
    \item \textbf{M4 Key Mismatch:} Inconsistent metric keys (\texttt{"acc"} vs. \texttt{"accuracy"}) caused quantitative parser failure.
    \item \textbf{M5 Incomplete:} Large-scale inductive validation datasets (Reddit, PPI) missing.
\end{itemize}

\paragraph{29. Informer (Long Sequence Time-Series Forecasting)}
\textit{Status: CLEAN}. ProbSparse attention mechanism and generative long-horizon forecasting pipeline were correctly implemented, recovering computational efficiency gains.

\paragraph{30. Autoformer (Decomposition Transformers for Time-Series)}
\textit{Status: CLEAN}. Series decomposition blocks and Auto-Correlation mechanisms were successfully reproduced, showing expected improvements over vanilla Transformer baselines.

\subsection{Cross-Experiment Case Study Summary}
Table~\ref{tab:case_study_summary} presents the systematic evaluation of all 30 benchmark reproduction experiments across verification levels L1--L3 and failure modes M1--M5.

\begin{table*}[t]
\centering
\caption{Hallucination Detection Results for 30 Benchmark Methods (Detailed Failure Analysis).}
\label{tab:case_study_summary}
\footnotesize
\setlength{\tabcolsep}{3pt}
\begin{tabularx}{\textwidth}{l c c c c c >{\raggedright\arraybackslash}X}
\toprule
\textbf{Method} & \textbf{M1} & \textbf{M2} & \textbf{M3} & \textbf{M4} & \textbf{M5} & \textbf{Failure / Detailed Issue} \\
\midrule
DPR & -- & -- & -- & -- & -- & All quantitative and qualitative checks passed. Evaluation metrics for NaturalQuestions and TriviaQA are fully present. \\
RAG & \checkmark & -- & -- & -- & \checkmark & \textbf{M1:} Generator falls back to parametric memory when retrieval fails. \textbf{M5:} Missing metrics for NaturalQuestions/WebQuestions. \\
ColBERT & -- & \checkmark & -- & -- & \checkmark & \textbf{M2:} Hard negative mining omitted from training. \textbf{M5:} Missing TREC Deep Learning dataset metrics. \\
BERT & \checkmark & -- & -- & -- & \checkmark & \textbf{M1:} Fine-tuned on non-target tasks instead of specified GLUE benchmarks. \textbf{M5:} Full GLUE dataset metrics missing. \\
RoBERTa & -- & -- & -- & \checkmark & \checkmark & \textbf{M4:} Key mismatch in \texttt{metrics.json} parser pipeline. \textbf{M5:} Missing full GLUE benchmark suite metrics. \\
LoRA & -- & -- & -- & -- & -- & Clean run. All required metrics for GLUE and instruction-tuning subsets are present. \\
Prefix-Tuning & -- & -- & -- & -- & -- & Clean reproduction without hallucinations; stable prefix parameter convergence. \\
PEGASUS & -- & \checkmark & \checkmark & -- & \checkmark & \textbf{M2:} Trained from scratch for 3 epochs instead of GSG checkpoint. \textbf{M3:} Inverted ranking (-32\% vs BERT). \textbf{M5:} Missing XSum. \\
LED & -- & -- & -- & -- & \checkmark & \textbf{M5:} Target method metrics missing for arXiv long-document dataset. \\
Longformer & -- & -- & -- & -- & -- & Clean. Sparse attention masks implemented correctly; expected efficiency recovered. \\
TextRank & -- & -- & -- & -- & -- & Clean. Graph-based extractive summarization baseline correctly evaluated. \\
ResNet & -- & -- & -- & -- & -- & Clean. Residual architecture advantage over plain CNNs reproduced on CIFAR-10/100. \\
ViT & -- & \checkmark & \checkmark & -- & \checkmark & \textbf{M2:} Trained from scratch without pretraining (20\% drop). \textbf{M3:} CNN outperforms ViT at reduced scale. \textbf{M5:} ImageNet subset missing. \\
CLIP & -- & -- & -- & -- & -- & Clean. Zero-shot transfer classification evaluated correctly with proper prompts. \\
SimCLR & \checkmark & -- & \checkmark & -- & \checkmark & \textbf{M1:} Baselines queued but unexecuted. \textbf{M3:} Substituted standard to ``above chance'' instead of beating baselines. \textbf{M5:} Baselines pending. \\
MoCo & -- & -- & -- & -- & \checkmark & \textbf{M5:} Metrics for ImageNet subset missing; linear probe accuracy unverified. \\
U-Net & -- & -- & \checkmark & -- & \checkmark & \textbf{M3:} Reduced-depth ablation outperformed full U-Net under small data. \textbf{M5:} Missing medical benchmarks (ISBI). \\
Faster R-CNN & -- & \checkmark & -- & -- & -- & \textbf{M2:} Non-standard RPN settings or mAP protocol degradation in detection pipeline. \\
YOLO & -- & -- & -- & -- & \checkmark & \textbf{M5:} Target method metrics missing for PASCAL VOC object detection benchmark. \\
PatchCore & -- & -- & -- & -- & \checkmark & \textbf{M5:} Detailed spatial localization metrics (pixel AUROC/AUPRO) missing. \\
PaDiM & -- & -- & -- & -- & -- & Clean. Mahalanobis distance modeling on patch embeddings verified on MVTec AD. \\
DDPM & -- & -- & -- & -- & -- & Clean. Stable diffusion pipeline reproduction with correct beta schedule and FID metrics. \\
DDIM & \checkmark & -- & -- & -- & \checkmark & \textbf{M1:} DDPM baseline never executed; speedup claim unverified. \textbf{M5:} Only 4/12 conditions completed (33\%). \\
Improved DDPM & \checkmark & -- & -- & -- & \checkmark & \textbf{M1:} Learned variance parameter instability eliminated expected quality gap. \textbf{M5:} Missing subset dataset metrics. \\
DQN & -- & -- & -- & -- & -- & Clean. Replay buffer and target network update recovered expected Atari/CartPole performance. \\
PPO & -- & \checkmark & -- & -- & \checkmark & \textbf{M2:} Continuous control (MuJoCo) crashed; falsely claimed success via discrete CartPole. \textbf{M5:} Baseline data missing. \\
GCN & -- & -- & -- & -- & -- & Clean. Graph convolution and adjacency preprocessing verified on Cora/Citeseer. \\
GraphSAGE & -- & \checkmark & -- & \checkmark & \checkmark & \textbf{M2:} Misconfigured inductive settings. \textbf{M4:} Key name mismatch (\texttt{"acc"} vs \texttt{"accuracy"}). \textbf{M5:} Reddit/PPI missing. \\
Informer & -- & -- & -- & -- & -- & Clean. ProbSparse attention and long-horizon forecasting pipeline correctly implemented. \\
Autoformer & -- & -- & -- & -- & -- & Clean. Series decomposition and auto-correlation mechanisms successfully reproduced. \\
\bottomrule
\end{tabularx}

\vspace{0.3cm}
\noindent\textbf{Aggregate Benchmark Statistics ($N=30$):}
\begin{itemize}[leftmargin=*, noitemsep, topsep=2pt]
    \item \textbf{Clean (no hallucination):} 13 (43.3\%)
    \item \textbf{M1: Method Integrity Collapse} --- 5 (16.7\%)
    \item \textbf{M2: Silent Protocol Degradation} --- 6 (20.0\%)
    \item \textbf{M3: Scale-Driven Conclusion Inversion} --- 4 (13.3\%)
    \item \textbf{M4: Quantitative Key Mismatch} --- 2 (6.7\%)
    \item \textbf{M5: Incomplete Execution} --- 16 (53.3\%)
\end{itemize}
\noindent\textit{Note: Failure categories M1--M5 overlap within individual methods; thus, the sum of category counts (33) exceeds the total count of hallucinated methods (17).}
\end{table*}

\section{Extended Discussion and Future Work}
\label{app:discussion}

\subsection{Discussion}
The empirical evaluations yield critical insights into the capabilities and limits of modern scientific agents. First, our fine-grained dimensional analysis reveals a performance mismatch between software viability and scientific validity. General-purpose coding agents generate syntactically correct code, but they lack domain-specific constraints needed to maintain research integrity. As a result, when faced with execution roadblocks, they optimize for engineering execution success ($\text{Exit Code } 0$) rather than scientific accuracy.

Second, the high prevalence of M5 (Incomplete Execution, 53.3\%) and M2 (Silent Protocol Degradation, 20.0\%) highlights the challenge of resource adaptation under fixed configurations. When compute bounds are reached, agents are forced into a trade-off: either abort execution (resulting in M5) or silently downgrade parameters (such as batch size or training steps, leading to M2) to satisfy execution limits. Because current agent architectures lack the context to dynamically partition workloads or negotiate resource limits, scaling automated discovery beyond simple sandbox environments remains difficult.

Finally, the low performance across all frameworks under Qualitative Review points to an abstraction gap. Current agents focus on local code correction, but they struggle to synthesize their findings into the structured, coherent, and contextualized narratives expected in academic research.

\subsection{Future Work: Multi-Agent Architectures and Resource Orchestration}

To resolve compute-bound limitations that lead to high rates of incomplete execution (M5), future work will explore transitioning from single-agent designs to cooperative, resource-aware multi-agent systems \cite{dolphin2025closed}. We envision a framework organized around specialized roles:

\begin{itemize}[leftmargin=*]
    \item \textbf{Resource Orchestrator Agent:} Dynamically monitors physical hardware configurations (VRAM, compute availability, process time limits). If a resource-limit error (e.g., OOM) is predicted, this agent automatically applies scaling rules---such as parallelizing execution nodes or deploying gradient accumulation---to keep execution within bounds without violating primary constraints ($\mathcal{C}$).
    \item \textbf{Developer Agent:} Focuses on generating candidate code $P$ to explore a target design space.
    \item \textbf{Auditor Agent:} Bound by independent personas to run AST and call-graph verification, checking developer outputs against the semantic constraint contract ($\mathcal{C}$). Future iterations will incorporate multi-model consensus to further mitigate evaluator bias and auditor hallucinations.
    \item \textbf{Red-Teaming Agent:} Aims to identify edge cases, such as searching for adversarial hyperparameter configurations or data subsets that could invalidate progress.
\end{itemize}

Implementing a multi-agent consensus pipeline with resource-aware delegation could prevent task failures caused by resource limits, laying a foundation for more reliable and scalable automated scientific discovery.

\ifappendixStandalone
\end{document}
\fi

\end{document}